\documentclass[prd,twocolumn,nofootinbib,showpacs,superscriptaddress]{revtex4-1}

\usepackage{amsfonts}
\usepackage{amsmath}
\usepackage{amssymb}
\usepackage{bm}
\usepackage{dcolumn}
\usepackage{graphicx}
\usepackage{graphics}
\usepackage[latin1]{inputenc}
\usepackage{latexsym}
\usepackage{rotating}
\usepackage{xspace} 
\usepackage[usenames]{color}
\usepackage{mathrsfs}
\usepackage{subfigure}

\usepackage{ulem}
\usepackage{pdfpages}
\usepackage{pgffor}

\makeatletter
\AtBeginDocument{\let\LS@rot\@undefined}
\makeatother

\definecolor {darkgreen}{rgb}{0.2,0.7,0.2}

\newcommand\be{\begin{equation}}
\newcommand\ba{\begin{eqnarray}}
\newcommand\ee{\end{equation}}
\newcommand\ea{\end{eqnarray}}

\newcommand\bw{\begin{widetext}}
\newcommand\ew{\end{widetext}}

\newcommand{\nn}{\nonumber}

\newcommand{\ppE}{{\mbox{\tiny ppE}}}

\newcommand{\GW}{{\mbox{\tiny GW}}}

\newcommand{\GR}{{\mbox{\tiny GR}}}

\begin{document}


# Erratum: A Parametrized post-Einsteinian Framework for Gravitational Wave Bursts

Nicholas Loutrel,[1] Nicolás Yunes,[1] and Frans Pretorius[2]

[1]*eXtreme Gravity Institute, Department of Physics,*
*Montana State University, Bozeman, MT 59717, USA.*
[2]*Department of Physics, Princeton University, Princeton, NJ 08544 USA.*

(Dated: September 26, 2017)

In [1], we developed a parameterized post-Einsteinian (ppE) model for the evolution of gravitational wave (GW) bursts from highly eccentric binary systems in time-frequency space. We here correct a few mistakes in some ingredients of the model.

## I. KEPLER PROBLEM IN THE PPE FORMALISM

As part of the computation of a ppE burst model, we considered an effective one body description for binary systems with generic modifications to the Newtonian potential. We claimed that the Newtonian potential within a generic modified theory of gravity can be written as shown in Eq. (B1) of [1], where the corrections scale as $M/r_p$, with $M$ the total mass of the binary and $r_p$ the pericenter distance of the binary. While the derivations in Appendix B of [1] are self-consistent, this scaling of the corrections is not suitable for some modified theories. We here generalize the results of this appendix to allow the model to apply to a wider set of theories.

For equatorial orbits, the corrections to the Newtonian kinetic energy and gravitational potential within a modified theory of gravity may be written as

$$\delta T = \frac{1}{2}\mu \left[ r^2\dot{\phi}^2 \tilde{\beta} \left(\frac{M}{r}\right)^{\tilde{b}} + \dot{r}^2 \tilde{\gamma}\left(\frac{M}{r}\right)^{\tilde{c}} \right], \quad (1)$$

$$\delta U = -\frac{\mu M}{r}\left[ \tilde{\alpha}\left(\frac{M}{r}\right)^{\tilde{a}} \right], \quad (2)$$

where $\mu$ is the reduce mass of the binary, $(\tilde{\alpha}, \tilde{\beta}, \tilde{\gamma})$ are amplitude parameters that depend on the coupling constants of the theory, and $(\tilde{a}, \tilde{b}, \tilde{c}) \in \mathbb{R}$ that control the post-Newtonian (PN) order of the modifications. This expression is similar to Eq. (B1) in [1], except that here we include an additional term proportional to $\dot{r}$, where $(\tilde{\gamma}, \tilde{c})$ are not related to the modification to the energy flux due to GW emission given in Eq. (29) of [1]. Moreover, the terms proportional to $(\tilde{\alpha}, \tilde{\beta}, \tilde{\gamma})$ depend on $(M/r)$ instead of $(M/r_p)$ as in Eq. (B1) in [1]. Equations (1) and (2) are indeed the form of the corrections in EDGB gravity.[1]

With this potential, we now compute the corrections to orbital quantities necessary to construct the ppE burst model, namely the orbital energy, angular momentum, period, and pericenter velocity. Let us begin by considering the Lagrangian for the binary within an effective one-body formalism. At leading (Newtonian) PN order, the Lagrangian for equatorial orbits may be written as $\mathcal{L} = T - U$, with $T = (1/2)\mu(\dot{r}^2 + r^2\dot{\phi}^2) + \delta T$ and $U = -\mu M/r + \delta U$. This Lagrangian admits two conserved quantities associated with the orbital energy and angular momentum, specifically

$$E = \frac{1}{2}\mu\dot{r}^2\left[1+\tilde{\gamma}\left(\frac{M}{r}\right)^{\tilde{c}}\right] + \frac{L^2}{2\mu r^2}\left[1-\tilde{\beta}\left(\frac{M}{r}\right)^{\tilde{b}}\right]$$
$$-\frac{\mu M}{r}\left[1+\tilde{\alpha}\left(\frac{M}{r}\right)^{\tilde{a}}\right] \quad (3)$$

$$L = \mu r^2\dot{\phi}\left[1+\tilde{\beta}\left(\frac{M}{r}\right)^{\tilde{b}}\right]. \quad (4)$$

By studying the turning points ($\dot{r}=0$) of Eq. (3), and writing $E = E_{\rm N} + \delta E$ and $L = L_{\rm N} + \delta L$, we find

$$\frac{\delta E}{E_{\rm N}} = -\frac{1}{2}\tilde{\alpha}\left(\frac{M}{r_p}\right)^{\tilde{a}}\mathcal{G}(e;\tilde{a},0,1) - \frac{1}{4}\tilde{\beta}\left(\frac{M}{r_p}\right)^{\tilde{b}}\mathcal{G}(e;\tilde{b},1,1), \quad (5)$$

$$\frac{\delta L}{L_{\rm N}} = \frac{1}{4}\tilde{\alpha}\left(\frac{M}{r_p}\right)^{\tilde{a}}\mathcal{G}(e;\tilde{a},1,0) + \frac{1}{8}\tilde{\beta}\left(\frac{M}{r_p}\right)^{\tilde{b}}\mathcal{G}(e;\tilde{b},2,0), \quad (6)$$

with the Newtonian energy and angular momentum $(E_{\rm N}, L_{\rm N})$ given in Eqs. (5.2) and (5.3) in [2], and

$$\mathcal{G}(e;q,m,n) = \frac{(1+e)^m(1-e)^n}{e}\left[1-\left(\frac{1-e}{1+e}\right)^{q+m-n}\right]. \quad (7)$$

These deformations can be related to the generic deformation for the energy and angular momentum given in

---

[1] These are not the only terms that appear in EDGB. Due to the nonlinearity of the point particle Lagrangian, higher order terms in the velocity can couple to the modifications to the metric. Here, we are only concerned with the leading order terms, i.e. those that scale as $v^2 \times (\tilde{\alpha}, \tilde{\beta}, \tilde{\gamma})$.

Eqs. (27) and (28) in [1], specifically

$$\alpha \left(\frac{M}{r_p}\right)^{\bar{a}} = \text{LO}\left[\frac{\delta E}{E_N}\right] \qquad \beta \left(\frac{M}{r_p}\right)^{\bar{b}} = \text{LO}\left[\frac{\delta L}{L_N}\right] \quad (8)$$

where LO stands for the leading-order term in a PN expansion.

Now, let us consider the orbital period within our ppE formalism. In order to do this, it is useful to parametrize the radius of the orbit as

$$r = \frac{p}{1 + e \cos\psi}, \quad (9)$$

where $p = r_p(1 + e)$ is the semi-latus rectum of the orbit and $\psi$ increases monotonically by $2\pi$ from one pericenter passage to the next. The reason for this parametrization is that, in general, any deformation to the Newtonian potential will cause the orbits to precess [3], and as a result, the orbit will return to pericenter when $\phi = 2\pi + \mathcal{O}(M/r_p)$. Using $\psi$ to parameterize the orbit allows us to avoid complications from precession when computing integrals of the orbital dynamics.

To obtain the evolution equation for $\psi$, we insert the radial function of Eq. (9) into Eq. (3) and solve for $\dot\psi$. The orbital period can then be computed by integrating $\dot\psi^{-1}$ between pericenter passages. Following this procedure, we obtain $T_\text{orb} = T_\text{orb}^N + \delta T_\text{orb}$, where $T_\text{orb}^N$ is given by Eq. (16) in [1] and

$$\frac{\delta T_\text{orb}}{T_\text{orb}^N} = \frac{1}{8\pi}\tilde\alpha\left(\frac{M}{r_p}\right)^{\tilde a}\mathcal{P}_1(e;\tilde a,1) + \frac{1}{16\pi}\tilde\beta\left(\frac{M}{r_p}\right)^{\tilde b}\mathcal{P}_1(e;\tilde b,2)$$
$$+ \frac{1}{4\pi}\tilde\gamma\left(\frac{M}{r_p}\right)^{\tilde c}\mathcal{P}_2(e), \quad (10)$$

with

$$\mathcal{P}_1(e;p,n) = \frac{(1-e^2)^{3/2}}{e^2(1+e)^p}\int_0^{2\pi}\frac{d\psi}{s^2(1+e\,c)^2}$$
$$\times\left[(1-e)^{p+n}(1-c)(2+e+e\,c)\right.$$
$$+(1+e)^{p+n}(1+c)(2-e+e\,c)$$
$$\left.-4(1+e\,c)^{p+n}\right], \quad (11)$$

$$\mathcal{P}_2(e) = \frac{(1-e^2)^{3/2}}{(1+e)^{\tilde c}}\int_0^{2\pi}d\psi(1+e\,c)^{\tilde c-2}, \quad (12)$$

and $(c,s) = (\cos\psi,\sin\psi)$.

Finally, consider the pericenter velocity. Pericenter corresponds to one of two points with $\dot r = 0$, so the velocity at this point is given by $v_p = r\dot\phi$ evaluated at $\psi = 2n\pi$, where $n \in \mathbb{Z}$. Evaluating this, we find $v_p = v_p^N + \delta v_p$, where $v_p^N = [M(1+e)/r_p]^{1/2}$, and

$$\frac{\delta v_p}{v_p^N} = \frac{1}{4}\tilde\alpha\left(\frac{M}{r_p}\right)^{\tilde a}\mathcal{V}_\alpha(e) - \frac{1}{8}\tilde\beta\left(\frac{M}{r_p}\right)^{\tilde b}\mathcal{V}_\beta(e), \quad (13)$$

with

$$\mathcal{V}_\alpha(e) = 1 + \left(\frac{1-e}{1+e}\right)^{\tilde a} + \frac{1}{e}\left[1 - \left(\frac{1-e}{1+e}\right)^{\tilde a}\right], \quad (14)$$

$$\mathcal{V}_\beta(e) = 4 + \left[2 - \left(\frac{1+e^2}{e}\right)\right]\left[1 - \left(\frac{1-e}{1+e}\right)^{\tilde b}\right]. \quad (15)$$

This completes the generalization of Appendix B in [1] and provides the necessary ingredients for the conservative sector of the ppE formalism.

## II. PPE BURST MODEL

How do the results of the previous section affect the ppE burst model developed in Sec. III of [1]? The generic form presented in Eqs. (1)-(6) of [1] are unchanged, only the mappings between ppE deformations given in Eqs. (27)-(30) in [1] and the ppE parameters that define the deformations to the burst model, specifically $\delta\lambda^a = (\alpha_\text{ppE},\beta_\text{ppE},\gamma_\text{ppE},\delta_\text{ppE})$ and $\delta\ell^a = (\bar a_\text{ppE},\bar b_\text{ppE},\bar c_\text{ppE},\bar d_\text{ppE})$.

The results of the previous section suggest two things. First, that there are a more fundamental set of ppE parameters, namely $(\tilde\alpha,\tilde\beta,\tilde\gamma)$ and $(\tilde a,\tilde b,\tilde c)$, that define the conservative orbital dynamics than the corrections used in [1] for the energy and angular momentum, specifically $(\alpha,\beta)$ and $(\bar a,\bar b)$. Thus, $\delta\lambda^a$ and $\delta\ell^a$ are more naturally mapped to $(\tilde\alpha,\tilde\beta,\tilde\gamma)$ and $(\tilde a,\tilde b,\tilde c)$.

Second, the only two quantities that are altered from [1] are the orbital period and pericenter velocity. In the burst model, the time centroids of the bursts are related by the orbital period, while the frequency centroids and widths of the bursts are characterized by the pericenter passage timescale, which in turn depends on the pericenter velocity through $\tau = r_p/v_p$. Thus the only quantities that change are the amplitude coefficients $(\alpha_\text{ppE},\beta_\text{ppE})$ and the exponent coefficients $(\bar a_\text{ppE},\bar b_\text{ppE})$ in the burst model.

With these considerations, the corrected mappings are

$$\alpha_\text{ppE}\left(\frac{M}{r_p}\right)^{\bar a_\text{ppE}} = \text{LO}\left[\frac{\delta T_\text{orb}}{T_\text{orb}^N}\right], \quad (16)$$

$$\beta_\text{ppE}\left(\frac{M}{r_p}\right)^{\bar b_\text{ppE}} = \text{LO}\left[\frac{\delta v_p}{v_p^N}\right]. \quad (17)$$

where $\delta T_\text{orb}$ and $\delta v_p$ are given in Eqs. (10) and (13), respectively. These correct Eqs. (44) and (45) in [1]. The mappings for $(\gamma_\text{ppE},\bar c_\text{ppE})$ and $(\delta_\text{ppE},\bar d_\text{ppE})$ are still given by Eqs. (46) and (47) in [1]. The burst model is now fully determined within this updated ppE formalism.

## III. BURSTS IN EDGB GRAVITY

As an application of the ppE burst model, we considered GW bursts in Einstein-Dilaton-Gauss-Bonnet

2

(EDGB) gravity in [1]. With the results of Sec. I, we here provide a corrected calculation for EDGB.

We begin by finding the leading-order contributions to the Newtonian gravitational potential in EDGB. To do this, we consider a point particle moving in a background spacetime of a non-rotating black hole in EDGB, which is given by

$$ds^2 = -f(r)\left[1 + h(r)\right]dt^2 + \frac{1}{f(r)}\left[1 + k(r)\right]dr^2 + r^2 d\Omega^2 \quad (18)$$

where $f(r) = 1 - 2M/r$ is the Schwarzschild factor, $h(r) = [\zeta/3f(r)](M/r)^3 \tilde{h}(r)$, $k(r) = -[\zeta/f(r)](M/r)^2 \tilde{k}(r)$, $\tilde{h}(r)$ and $\tilde{k}(r)$ are given by Eqs. (10) and (11) in [4], and $\zeta = \xi/M^4$ is the dimensionless coupling constant of the theory. The Lagrangian for the test particle is given by $\mathcal{L} = \mu\sqrt{-g_{\rho\sigma}v^\rho v^\sigma}$, where $v^\rho = (1, dx^j/dt)$ and we have taken the mass of the particle to be the reduced mass of the binary considered in Sec. I. By performing a PN expansion of this Lagrangian, we find the corrections to the potential and kinetic energy to lowest order in $\zeta$ and $M/r$ to be

$$\delta U = \frac{\zeta}{6}\mu\left(\frac{M}{r}\right)^3, \qquad \delta T = -\frac{\zeta}{2}\mu \dot{r}^2 \left(\frac{M}{r}\right)^2. \quad (19)$$

Matching this to Eq. (2), we have

$$\tilde{\alpha} = -\frac{\zeta}{6}, \qquad \tilde{a} = 2, \quad (20)$$
$$\tilde{\gamma} = -\zeta, \qquad \tilde{c} = 2. \quad (21)$$

We do not consider the corrections to the potential that scale as $\dot{\phi}^2$ for two reasons: these contributions are higher PN order, and they do not admit the simple form presented in Eq. (1) due to the nonlinearity of the Lagrangian. Neglecting these terms does not affect the leading-order results presented in this section.

We may now directly apply the results of Sec. I to EDGB gravity. Below we provide expressions for the orbital energy, period, and pericenter velocity for eccentric binaries in EDGB:

$$\frac{\delta E_{\rm EDGB}}{E_{\rm N}} = \frac{\zeta}{6}\left(\frac{1-e}{1+e}\right)\left(\frac{M}{r_p}\right)^2, \quad (22)$$

$$\frac{\delta T_{\rm orb}^{\rm EDGB}}{T_{\rm orb}^{\rm N}} = -\frac{\zeta}{4}\left(\frac{1-e}{1+e}\right)\left(1 + 2\sqrt{1-e^2}\right)\left(\frac{M}{r_p}\right)^2, \quad (23)$$

$$\frac{\delta v_p^{\rm EDGB}}{v_p^{\rm N}} = -\frac{\zeta}{12}\left[\frac{3+e^2}{(1+e)^2}\right]\left(\frac{M}{r_p}\right)^2. \quad (24)$$

which corrects the energy displayed in Eq. (50) of [1]. This agrees with the circular limit in [4] and we have independently verified the above results using the methods therein. The angular momentum remains the same as reported in Eq. (51) of [1]. Thus, the correct ppE parameters for the burst model in EDGB are

$$\alpha_{\rm ppE} = -\frac{\zeta}{8}\delta e + \mathcal{O}(\delta e^2), \qquad \bar{a}_{\rm ppE} = 2, \quad (25)$$
$$\beta_{\rm ppE} = -\frac{\zeta}{12} + \mathcal{O}(\delta e), \qquad \bar{b}_{\rm ppE} = 2, \quad (26)$$

where we have expanded about $\delta e = 1 - e \ll 1$. Notice that the $\bar{b}_{\rm ppE}$ parameter is 2 and not 3 as reported below Eq. (55) in [1], and that $\alpha_{\rm ppE}$ is not zero as reported in Eq. (56) of that same paper. The updated time-frequency mapping of the bursts is then given by

$$t_i^{\rm EDGB} = t_{i-1} + \Delta t_{(i,i-1)}^{\rm GR}\left[1 - \frac{\zeta}{8}\delta e_i \left(\frac{M}{r_{p,i}}\right)^2\right] \quad (27)$$

$$f_i^{\rm EDGB} = f_i^{\rm GR}\left[1 - \frac{\zeta}{12}\left(\frac{M}{r_{p,i-1}}\right)^2\right] \quad (28)$$

$$\delta t_i^{\rm EDGB} = \delta t_i^{\rm GR}\left[1 + \frac{\zeta}{12}\left(\frac{M}{r_{p,i-1}}\right)^2\right] \quad (29)$$

$$\delta f_i^{\rm EDGB} = \delta f_i^{\rm GR}\left[1 - \frac{\zeta}{12}\left(\frac{M}{r_{p,i-1}}\right)^2\right] \quad (30)$$

which corrects Eqs. (57)-(60) in [1]. The remainder of the burst model in EDGB is unchanged. It is worth noting that these corrections to the EDGB burst model do not alter the projected constraint that can be placed on $\xi$, which is given by Eq. (93) in [1]. This result comes from considering the lowest PN order correction to the burst model, corresponding to corrections to the GW energy and angular momentum fluxes due to dipole radiation.


\clearpage
\title{A Parametrized post-Einsteinian Framework for Gravitational Wave Bursts}

\author{Nicholas Loutrel}
\affiliation{Department of Physics, Montana State University, Bozeman, MT 59717, USA.}

\author{Nicol\'as Yunes}
\affiliation{Department of Physics, Montana State University, Bozeman, MT 59717, USA.}

\author{Frans Pretorius}
\affiliation{Department of Physics, Princeton University, Princeton, NJ 08544 USA.}

\date{\today}

\begin{abstract} 

The population of stellar-mass, compact object binaries that merge with non-negligible
eccentricity may be large enough to motivate searches with
ground-based gravitational wave detectors. Such events could be exceptional
laboratories to test General Relativity in the dynamical, strong-field regime, as
a larger fraction of the energy is emitted at high-velocities, compared to quasi-circular
inspirals. A serious obstacle here, however, is the challenge of computing theoretical waveforms 
for eccentric systems with the requisite accuracy for use in a matched-filter search. 
The corresponding waveforms are more a sequence
of concentrated \emph{bursts} of energy emitted near periapse than a continuous waveform. 
Based on this, an alternative approach, stacking excess power over the set of
time-frequency tiles coincident with the bursts, was recently suggested as a more practical (though
sub-optimal) detection strategy. The leading-order ``observable'' that would be inferred
from such a detection would be a sequence of discrete numbers characterizing the position
and size of each time-frequency tile. In General Relativity, this (possibly large) sequence of 
numbers is uniquely determined by the small set of parameters describing the binary at formation. 
In this work, following the spirit of the parameterized post-Einsteinian framework developed for
quasi-circular inspiral, we propose a simple, parameterized deformation of the baseline
general relativistic burst algorithm for eccentric inspiral events that would 
allow for model-independent tests of Einstein's theory in this high-velocity, strong-field regime.

\end{abstract}


\maketitle

\section{Introduction}

The dynamical and non-linear, strong-field regime of General Relativity (GR) is essentially unconstrained
by observations and experiment~\cite{will-living,lrr-2013-9}. One characterization of this regime
is a scenario where a source of large spacetime curvature experiences sufficiently rapid
acceleration (as measured by a distant inertial observer) to produce gravitational 
wave (GW) emission approaching the Planck luminosity $L_p=c^5/G$. 

Vacuum GR has no intrinsic scale, so one could argue that
there is nothing particularly relevant about $L_p$ compared to any other
unit-luminosity in some other system of units. However, if one considers
a particular GW emission process with an energy scale $E=M c^2$ (here $M$ is the
total rest mass of the binary) occurring within a region of size $R$ (here, the binary diameter), 
arguments based on the Hoop Conjecture~\cite{thorne_hoop} suggest this can only exceed $L_p$ if cosmic 
censorship is violated. Thus, in GR $L_p$ is a natural scale to expect the most radical
phenomena of the theory to occur. 

At luminosities of ${\cal{O}}(10^{-27})L_p$, the double pulsar PSR J0737-3039~\cite{kramer-double-pulsar} is to date the most luminous system with inferred GW emission, yet this alone is wholly inadequate to give confidence that GR's predictions 
hold all the way to the Planck scale. Fortunately, the next generation of ground based GW
detectors---LIGO~\cite{Abramovici:1992ah,Abbott:2007kv,Harry:2010zz,ligo}, KAGRA~\cite{Somiya:2011np,Aso:2013eba}, VIRGO~\cite{Giazotto:1988gw,Caron:1997hu,virgo}, GEO~\cite{Luck:1997hv,Willke:2002bs,Grote:2008zz,geo}---are expected 
to soon observe binary black hole, binary neutron star, and black hole-neutron star mergers; the final stages of these collisions 
are expected to reach $\sim 10^{-2} L_p$.

Compact object binaries that have large orbital eccentricity, while emitting GWs in the frequency band
of ground-based detectors, have long been dismissed as likely sources. This is 
because (i) GW emission is efficient at circularizing orbits with large pericenters, 
expected for the typical stellar binary progenitor systems, and (ii) more ``exotic''
channels of formation that could lead to small-pericenter/high-eccentricity systems
were thought to be sufficiently rare to be irrelevant for the up and coming generation of 
detectors. 

Although this may still be the prevailing view in the GW source modeling and detection communities, 
within the past several years more detailed studies of alternative formation channels suggest
eccentric merger rates may be high enough to make them
a plausible class of events to search for. Such mechanisms include dynamical capture during a close two-body
encounter in a dense stellar environment~\cite{2009MNRAS.395.2127O,lee2010}, 
a single-binary interaction in a similar environment~\cite{Samsing:2013kua}, and the Kozai-Lidov
mechanism in a hierarchical triple system~\cite{2003ApJ...598..419W,Kushnir:2013hpa,2013PhRvL.111f1106S,
2013arXiv1308.5682A,2013ApJ...773..187N,2014ApJ...781...45A} (for more information see the 
discussions in~\cite{2013PhRvD..87d3004E,up_and_coming}). 

\begin{figure}
\includegraphics[clip=true,width=9.5cm]{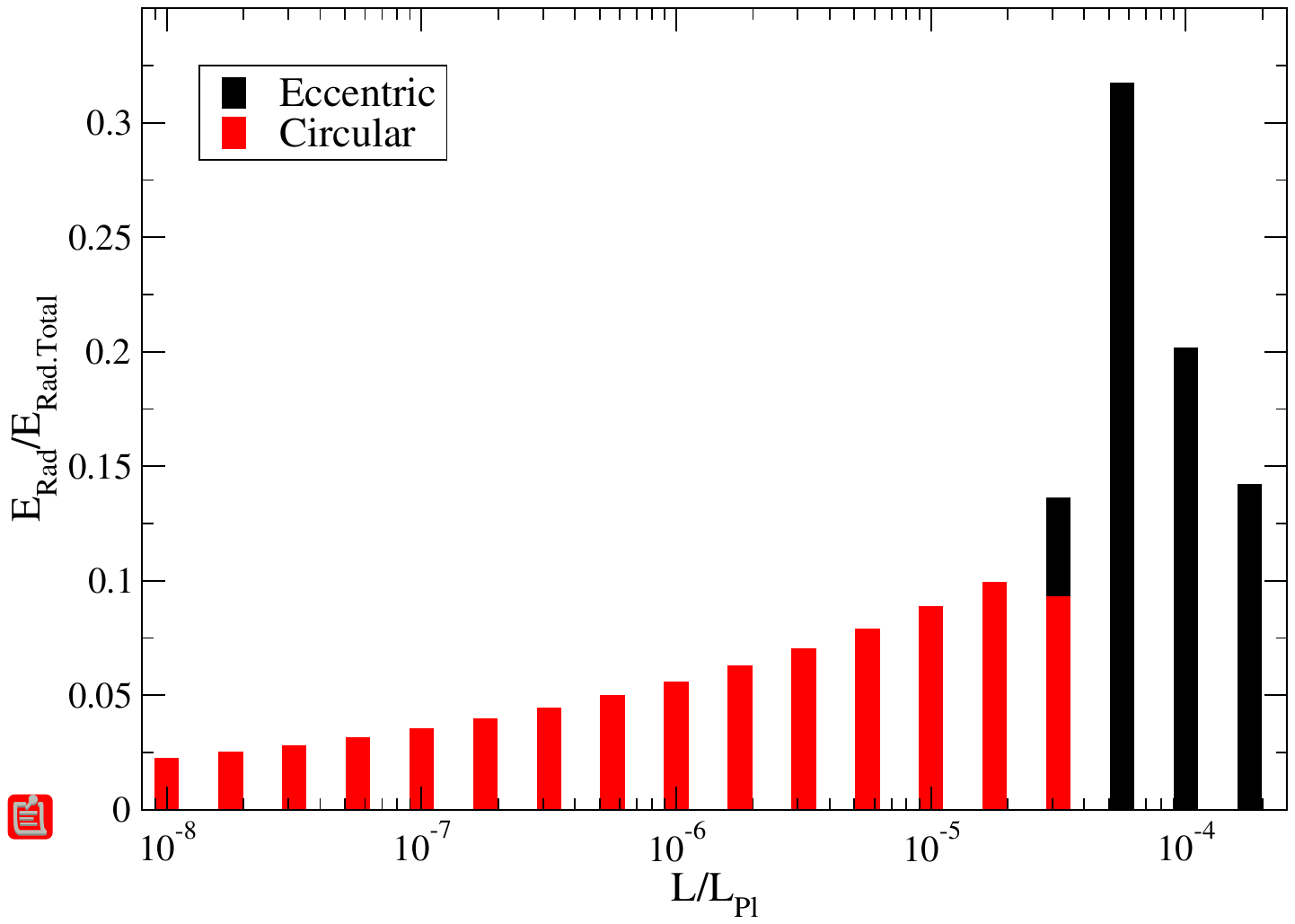}
\caption{\label{luminosity} Histogram illustrating the fraction of GW energy radiated as a function of the GW luminosity in Planck units for an eccentric (black) and a quasicircular (red) inspiral, computed using the Newtonian orbit plus quadrupole emission model discussed in the text. The eccentric inspiral starts at an initial pericenter of $8 M$ and eccentricity of $0.99$. The quasicircular inspiral starts at a separation of $800 M$, corresponding to the same initial semi-major axis as the eccentric inspiral. Both orbits are terminated at the innermost stable orbit at pericenter distance $r_{p} = (3 + e_{f})/(1 + e_{f}) 2 M$, with $e_f$ the ``final'' eccentricity before plunge. We have thus left out the GW energy emitted during plunge, merger and ringdown, which in both cases would contribute to the energy at luminosities of order $10^{-3}$ to $10^{-2} L_p$. Observe that while the quasi-circular inspiral emits radiation at all luminosities below this, the eccentric inspiral does not emit energy below $10^{-5} L_p$ for the system considered, concentrating all its power at higher luminosities.}
\end{figure}

High-eccentricity mergers may be ideal for 
testing GR in the dynamical strong field regime, as, other
parameters in the system being equal the peak GW luminosity achieved
in these systems is typically higher than in a quasi-circular inspiral. Furthermore, although the
integrated power radiated is comparable, more of it comes from the
high-luminosity regime in eccentric mergers---see for example Fig.~\ref{luminosity}.
However, no studies have yet been done that show eccentric mergers would in {\em practice}
be better than quasi-circular systems for constraining strong-field gravity,
in particular given some of the issues with templates discussed below.
One goal of this paper is therefore to lay a theoretical foundation to 
quantitatively begin to address this, and related questions.

Using GW observations of compact binary mergers, eccentric or not, to test 
GR or search for deviations from its predictions is not a straightforward endeavor. 
One of the problems lies with the reliance on theoretical templates of expected events
to go beyond mere detection and infer properties of the source. Any aspect
of a detected source not reflected in the template bank, whether due to modeling
deficiencies within GR, or problems with the theory itself, will lead
to a misidentification of the source as the best-fit member of the
template bank, and by inference the theory used to construct it. 

This problem of {\em fundamental theoretical bias} in the GW detection
endeavor was the motivation for the development of the parameterized 
post-Einsteinian (ppE) framework~\cite{PPE}.
The basic idea is to begin with a class of sources
(quasi-circular, compact binary inspirals in the case of ~\cite{PPE}) for which there
is decent evidence that GR predictions are sufficiently
accurate for some fraction of events to still be detected with GR
templates. Then, one considers a set of ``well-motivated'' plausible
alterations to the theory---additional energy emission channels,
GW polarizations, new conservative terms that appear
at high $v/c$ in an effective Hamiltonian description, examples
from specific alternative theories, etc.---and
how these alterations imprint on the eventual observable 
(the response function in the case of ~\cite{PPE}). 
Following this exercise, one proposes a parameterized 
set of deformations of the baseline GR observable, that capture 
the particular case studies considered for specific
values of the parameters, and reduce to the GR result in the appropriate limit. 
After a putative detection of a source with GR templates, a follow-up study
with ppE templates could then be performed to either {\em quantify}
the consistency of the event with GR as the underlying theory, or
give evidence for a correction to GR. Several studies have since
illustrated the efficacy of this approach 
(see e.g.~\cite{cornishsampson,Sampson:2013lpa,Sampson:2013jpa}).

The main goal of this paper is to introduce a ppE framework to allow
the use of GW observations of high-eccentricity, merger events to 
test GR. One approach would simply be to extend the original
ppE waveforms to now include eccentricity as a parameter.
The problem with this is that it is unlikely that high-eccentricity GR templates with
sufficient accuracy to use in matched-filtering searches will be available
by the time detectors begin their next observing runs. This is because high-eccentricity 
coalescences lead to waves that are emitted in a sequence of concentrated 
\emph{bursts}---a large emission of radiation in a small time-frequency window---each associated with
periapse passage. Even a small error in the waveform phase could lead to a huge
timing error for when the next burst occurs, leading, in turn, to a huge loss of signal-to-noise ratio (SNR).

To partially alleviate this problem, Ref.~\cite{2013PhRvD..87d3004E} 
proposed that, as an alternative to matched filtering,
a power stacking algorithm could be employed, where 
power in a set of time-frequency \emph{tiles} informed by a theoretical
model could be added together to search for a statistically significant
excess. Each tile is centered about one of the GW bursts. The advantage of power stacking
is that the model only needs to predict the waveform to a phase accuracy
of order one cycle over the time of emission in band, compared
to a small fraction of a cycle required by matched filtering. The disadvantage
is that this will not give an optimal SNR; roughly speaking, for an 
$N$ burst event, power stacking will achieve an SNR proportional to $N^{1/4}$
times that of a single burst, versus $N^{1/2}$ using matched-filtering 
(for a more detailed analysis see~\cite{up_and_coming}). 
\subsection{Executive Summary}

We here adapt our ppE framework to highly-eccentric binaries, assuming a power-stacking
detection strategy. We follow the same procedure as that
outlined above for the original, quasi-circular ppE framework, with the key difference 
that the observable is now the set of parameters characterizing the burst sequence. 
Each burst can be described by four numbers that specify the center, width and height
of a tile in the time-frequency plane, capturing a desired fraction
of the energy of the burst. In GR, the sequence of bursts is uniquely
characterized by the small set of parameters describing the ``initial
conditions'' of the binary; to leading, \emph{Newtonian} order in a weak field expansion, these
are the symmetric mass ratio $\eta$, the total mass $M$, the pericenter
of the first encounter $r_{p,0}$, and the initial orbital eccentricity $e_0$. 

A ppE burst-sequence for a given event is a parameterized deformation of the GR burst sequence. 
Na\"ively, this would require a ppE parameter for each
of the $4N$ parameters of an $N$-burst sequence. However, as with the original
ppE, we restrict the class of deviations to those that are well-motivated
in the sense described previously. Using a simplified model of the GR
sequence to allow for an analytic study, we show that only 8 ppE parameters
suffice to capture a wide class of deformations. Physically, these parameters correspond
to deformations in the binary's binding energy and angular momentum (i.e.~the conservative sector)
and their rate of change (i.e.~the dissipative sector). We then postulate that
this ppE deformation can be applied to an arbitrarily accurate baseline
GR burst sequence model.

To be more concrete, we propose the following {\em burst algorithm}, the first key ingredient of which describes the properties of the GW emission coming from the $i^{th}$ burst in the sequence:
\allowdisplaybreaks[4]
\begin{align}
\label{eq:ppEexact1-new}
t_{i}^{\rm ppE} &= t_{i-1} + \Delta t_{i,i-1}^{\rm GR}(r_{p,i}, e_{i}) \left[1 + \alpha_{\ppE} \left(\frac{M}{r_{p,i}}\right)^{\bar{a}_{\ppE}} \right]\,,
\\
\label{f-new}
f_{i}^{\rm ppE} &= f_{i}^{\rm GR}(r_{p,i}, e_{i}) 
\left[1 + \beta_{\ppE}  \left(\frac{M}{r_{p,i}}\right)^{\bar{b}_{\ppE}} \right]\,,
\\
\delta t_{i}^{\rm ppE} &= \delta t_{i}^{\rm GR}(r_{p,i}, e_{i}) 
\left[1 -  \beta_{\ppE}  \left(\frac{M}{r_{p,i}}\right)^{\bar{b}_{\ppE}} \right]\,,
\\
\label{eq:ppEexact4-new}
\delta f_{i}^{\rm ppE} &= \delta f_{i}^{\rm GR}(r_{p,i}, e_{i}) 
\left[1 +  \beta_{\ppE}  \left(\frac{M}{r_{p,i}}\right)^{\bar{b}_{\ppE}} \right]\,.
\end{align}
Here $(\alpha_{\ppE},\beta_{\ppE})$ are ppE amplitude parameters, and $(\bar{a}_{\ppE},\bar{b}_{\ppE})$ are ppE exponent parameters. Equations~\eqref{eq:ppEexact1-new}-\eqref{eq:ppEexact4-new} describe the centroid ($t_{i},f_{i}$) and width ($\delta t_{i},\delta f_{i}$) of the $i^{th}$ tile in time and frequency. The quantities with the GR superscripts denote the pure GR values in the limit $(\alpha_{\ppE},\beta_{\ppE})=(0,0)$, and $\Delta t_{i,i-1}^{\rm GR}(r_{p,i}, e_{i})$ is the function giving the period of the orbit preceding the $i^{th}$ burst in GR (see e.g.~Eqs.~\eqref{t2-exact}-\eqref{deltaf2-exact}).

The GR functions and ppE corrections in Eqs.~\eqref{eq:ppEexact1-new}-\eqref{eq:ppEexact4-new} depend on the pericenter $r_{p,i}$ and eccentricity $e_{i}$ of the corresponding orbit, however the orbit changes from one burst to the next. Thus the second key ingredient to the burst algorithm describes the mapping from the $(i-1)^{th}$ to the $i^{th}$ orbit: 
\begin{align}
\label{rp-new}
& \frac{r_{p,i}^{\ppE}(r_{p,i-1},e_{i-1})}{r_{p,i}^{\GR}(r_{p,i-1},e_{i-1})} = 1 + \gamma_{\ppE} \left(\frac{M}{r_{p,i-1}}\right)^{\bar{c}_{\ppE}}\,,
\\ 
&\frac{\Delta \delta e_{i,i-1}^{\ppE}(r_{p,i-1},e_{i-1})}{\Delta \delta e_{i,i-1}^{\GR}(r_{p,i-1},e_{i-1})} =  1 + \delta_{\ppE} \left(\frac{M}{r_{p,i-1}}\right)^{\bar{d}_{\ppE}}\,.
\label{eq:deltas-new}
\end{align}
Here $(\gamma_{\ppE},\delta_{\ppE})$ are ppE amplitude parameters, and $(\bar{c}_{\ppE},\bar{d}_{\ppE})$ are ppE exponent parameters, and we have defined $\Delta \delta e_{i,i-1}= e_{i-1} - e_{i}$. As before, the quantities with a GR superscript represent the mappings in GR (see e.g.~Eqs.~\eqref{rp2}-\eqref{e2}). 

We only need 4 ppE amplitude parameters and 4 ppE exponent parameters to characterize deformations to eccentric bursts. The number of parameters makes sense, since such dynamical encounters are really controlled by only 4 quantities to leading-order in a weak-field expansion: the binding energy, the orbital angular momentum, the energy flux and the angular momentum flux. Deforming each of these relations through polynomials of the form $\epsilon_{i} \; (M/r_{p})^{e_{i}}$, for amplitude coefficients $\epsilon_{i}$ and exponent coefficient $e_{i}$,  then leads to deformations to the GR burst sequence that can only depend on 4+4 parameters. We also here calculate how the ppE parameters of the burst sequence map to deformations to the binding energy, angular momentum and fluxes, i.e.~how $(\alpha_{\ppE},\beta_{\ppE},\gamma_{\ppE},\delta_{\ppE})$ and $(\bar{a}_{\ppE},\bar{b}_{\ppE},\bar{c}_{\ppE},\bar{d}_{\ppE})$ map to $(\epsilon_{1},\epsilon_{2},\epsilon_{3},\epsilon_{4})$ and $(e_{1},e_{2},e_{3},e_{4})$. 

Finally, we investigate two specific modified theories of gravity, develop burst algorithms for them and determine whether they can be mapped to the above ppE framework. In particular, we consider massless, Brans-Dicke (BD) theory~\cite{Brans:1961sx,TEGP} and Einstein-Dilaton-Gauss-Bonnet (EDGB) theory~\cite{yunesstein,quadratic}. Both of these modify the Einstein-Hilbert action through a dynamical scalar field, sourced by the matter stress-energy tensor and by a certain combination of quadratic curvature invariants respectively. Both of them lead to dipolar GW emission, which we calculate here for the first time for non-spinning, eccentric inspirals. Such emission leads to a modified burst algorithm, which we prove explicitly maps to the ppE deformations proposed above.  

We emphasize again that our goal here is to construct a ppE deformation
of eccentric burst sequences. Clearly, much follow-up 
work is needed to actually demonstrate that this is a useful
approach to constrain GR or search for new physics in the strong-field regime.
We also note that an approach like this may be useful to search for
finite body and equation of state effects in high-eccentricity mergers involving neutron stars.
Namely, here one would assume GR is correct, use the GR point-particle (black hole)
sequence as the baseline, and add parameterized deformations to capture
difficult-to-model physics, finite body effects, or uncertainties in the nuclear equation of state.
These topics, as discussed in the conclusion, are left for future work.

The remainder of this paper presents the details of the calculation that lead to the results described above. Section~\ref{sec:GR-modeling} reviews how to build a burst algorithm for eccentric encounters in GR. Section~\ref{sec:modeling-beyond-GR} rederives such an algorithm in theories that are parametrically deformed away from GR, discussing the ppE parameterization in more detail. Section~\ref{sec:burst-in-mod-gravity} derives burst algorithms for two example theories. Section~\ref{conclusions} concludes and points to future research. Henceforth, we use geometric units where $G = 1 = c$. 

\section{Analytic Modeling in GR}
\label{sec:GR-modeling}

In this section we construct an analytical burst algorithm within GR. We assume that either the binary is already in a high-eccentricity orbit, or it is in a hyperbolic or parabolic orbit that after the first close encounter becomes bound due to GW emission. We derive the mapping between the time-frequency tiles of each burst to leading post-Newtonian (PN) order in a weak-field/slow-motion expansion. As such, this will only be a proof of concept model to guide the development of the ppE extension; a more realistic, higher-PN order or a numerical model will be built in future work.

\subsection{Size of Tiles}
\label{sec:size-of-tiles}

Let us first consider how to choose the time-frequency size of the burst tiles.
Since most of the radiation occurs when the system is in the neighborhood of the point of closest approach, 
the temporal width is proportional to the characteristic GW time~\citep{1977ApJ...216..610T}:
\begin{equation}
\label{tauGW-def}
\tau_{\GW} \equiv \frac{{\rm pericenter} \; {\rm{distance}}}{{\rm{pericenter}}\;{\rm{velocity}}} = \frac{r_{p}^{3/2}}{[M (1 + e)]^{1/2}}\,
\end{equation}
where $M$ is the total mass of the binary, $e$ is the eccentricity of the orbit and we have mapped the two-body problem to an effective-one-body problem. The temporal width of the tile can then be written as 
\begin{equation}
\delta t = \xi_{t} \; \tau_{\GW}\,
\label{deltat}
\end{equation}
where $\xi_{t}$ is a proportionality constant of order unity.
To leading order, the characteristic frequency of GW emission on a single pericenter passage will be proportional to the reciprocal of the characteristic time,
\begin{equation}
\label{fGW-def}
f_{\GW} = \frac{1}{2 \pi \tau_{\GW}}\,.
\end{equation} 
This frequency is roughly where most of the signal power is emitted, as discussed in~\citep{1977ApJ...216..610T,2010PhRvD..82j7501B}. One should then center the tiles at this characteristic frequency and choose the tile width to be
\begin{equation}
\delta f = \xi_{f} \; f_{\GW}\,
\label{deltaf}
\end{equation}
with proportionality constant $\xi_{f}$ of order unity. 
The choice of proportionality constants $\xi_{t}$ and $\xi_{f}$ would be determined by a combination of modeling and data analysis considerations. For example, they should be sufficiently large that the tile overlaps
the majority of the GW energy of actual bursts given some estimate of modeling errors, though not so large as to cause an unacceptable false-alarm rate. For more discussion on these issues see~\cite{up_and_coming}.

\subsection{Mapping between Tiles}
To construct a simple, analytical model to map from one burst tile to the next, we approximate
the binary orbit following the initial encounter as a sequence of Newtonian ellipses, with
the parameters of each ellipse changing instantaneously at pericenter passage due to the emission of GWs,
calculated to leading order in $v/c$. 
This is a decent approximation while the eccentricity is large, so that little energy
is emitted away from pericenter, and while pericenter distances are large, so that higher order terms
in $v/c$ are negligible (some justification for these statements is given in Appendix A).

Of course, for the class of binaries that could be within reach
of ground based detectors, much of the observable GW emission will come from stages of the
evolution where these approximations are poor. Regardless, we emphasize again that the use
of these approximations is only to allow a simple analysis of non-GR corrections to the dynamics that
later motivate the form of the ppE deformations given in Eqs.~(\ref{eq:ppEexact1-new}-\ref{eq:deltas-new}). In
those expressions, the GR mapping between bursts can be calculated as 
accurately as possible, for example through full numerical relativity simulations. 
Thus, in Eqs.~(\ref{eq:ppEexact1-new}-\ref{eq:deltas-new}) we target GR deformations that continue to follow
the same parametric deviations as derived below using the less accurate GR model.

First, we characterize the orbit (open or closed) preceding the first close encounter 
by an initial pericenter distance $r_{p,0}$ and eccentricity $e_{0}$. This
orbit will have an energy $E_{0}$ and angular momentum $L_{0}$.
About the time of closest approach, a burst of radiation is emitted, changing the energy and 
angular momentum by $\Delta E = E_{1} - E_{0}$ and $\Delta L = L_{1} - L_{0}$, which to 
leading PN order is~\cite{Peters:1964zz,PetersMathews}:
\begin{align}
\label{DeltaE}
\frac{\Delta E}{M} &= - \frac{64 \pi}{5} \eta^{2} \left(\frac{M}{r_{p,0}}\right)^{7/2} \!\!\! \left(1 + \frac{73}{24} e_{0}^{2} + \frac{37}{96} e_{0}^{4}\right),
\\
\label{DeltaL}
\frac{\Delta L}{M^{2}} &= - \frac{64 \pi}{5} \eta^{2} \left(\frac{M}{r_{p,0}}\right)^{2} \left(1 + \frac{7}{8} e_{0}^{2}\right).
\end{align}
The symmetric mass ratio $\eta = \mu/M$, where $\mu = (m_{1} m_{2})/M$ is the reduced mass.
If the initial system is unbound ($e_0>1$), then we require the first encounter to be sufficiently
close that the system becomes bound, namely that $r_{p,0} < r_{p,\rm{bound}}$ with 
\begin{align}
\label{eq:rp-bound}
\frac{r_{p,\rm{bound}}}{M} &\equiv \left(\frac{128 \pi}{5}\right)^{2/5} \left(\frac{\eta}{e_{0}-1}\right)^{2/5} 
\left(1 + \frac{73}{24} e_{0}^{2} + \frac{37}{96} e_{0}^{4}\right)^{2/5}\,. 
\end{align} 

The emission of energy and angular momentum due to this zeroth-encounter will lead to a new $r_{p,1}$ and $e_{1}$
characterizing the first full orbit in the sequence. These parameters can be obtained by solving for $(r_{p,1},e_{1})$ in terms of $E_{1}$ and $L_{1}$. To leading PN order, the relations between these
parameters are given by~\cite{Peters:1964zz}
\begin{align}
\label{r-inv}
r_{p} &= \frac{M}{2 {\cal{E}} } \;  \left(1 -  \sqrt{1 - 2 {\cal{E}} {\cal{L}}^{2}  } \right)\,.
\\
\label{e-inv}
e &= \sqrt{1 - 2 {\cal{E}} {\cal{L}}^{2} }\,,
\end{align}
where we have introduced the reduced energy ${\cal{E}} \equiv - E/(M \eta)$ and angular momentum ${\cal{L}} \equiv L/(M^{2} \eta)$.  
The time to the next burst is then given by the orbital period $T_{\rm orb}$ of the ellipse with
$(r_p,e)=(r_{p,1},e_{1})$:
\begin{align}
\label{torb}
T_{\rm orb} &= 2 \pi \sqrt{\frac{a^{3}}{M}} = \frac{2 \pi r_{p}^{3/2}}{(1 - e)^{3/2} M^{1/2}}
= \frac{2 \pi r_{p}^{3/2}}{{\delta e}^{3/2} M^{1/2}} \,,
\end{align}
where  $a$ is the semi-major axis of the ellipse, and we have defined $\delta e \equiv 1 - e$. 

Above, we have explicitly described the evolution of the system from the initial encounter to the
next, though as long as $\delta e$ remains small and $r_p$ large, all subsequent orbits can be calculated
applying exactly the same formulas. Doing so, to leading order in $\delta e$ and $M/r_p$ we find
the following map from the parameters of the $(i-1)^{th}$ to the $i^{th}$ orbit :
\allowdisplaybreaks[4]
\begin{align}
\label{rp2}
r_{p,i}^{\GR}(r_{p,i-1},e_{i-1}) &= r_{p,i-1} \left\{1 - \frac{59 \sqrt{2} \pi}{24} \eta \left(\frac{M}{r_{p,i-1}}\right)^{5/2} 
\right. 
\nn \\
&\left.\times
\left[1 + \frac{121}{236} {\delta e}_{i-1} \right] 
\right.
\nn \\
&+ \left. {\cal{O}}\left[\left(\frac{M}{r_{p,i-1}}\right)^{7/2},\delta e_{i-1}^{2}\right] \right\}\,,
\\
\label{e2}
\Delta \delta e_{i,i-1}^{\GR}(r_{p,i-1},e_{i-1}) &\equiv \delta e_{i}^{\GR}(r_{p,i-1},e_{i-1}) - \delta e_{i-1} \nn\\
&= \frac{85 \sqrt{2} \pi}{12} \eta \left(\frac{M}{r_{p,i-1}}\right)^{5/2}
\nn \\
&\left(1 - \frac{543}{1700} \delta e_{i-1}\right) 
\nn \\
&+ {\cal{O}}\left[\left(\frac{M}{r_{p,i-1}}\right)^{7/2},\delta e_{i-1}^{2}\right].
\end{align}
Note that the above definition of $\Delta \delta e_{i,i-1}$ is equivalent to its previous one ($\Delta \delta e_{i,i-1} = e_{i-1} - e_{i}$). The time between the $(i-1)^{th}$ and $i^{th}$ burst is the period of the $i^{th}$ orbit
\begin{align}
\label{t2-exact}
\Delta t_{i,i-1} ^{\GR}(r_{p,i-1},e_{i-1}) =\frac{2 \pi [r_{p,i}^{\GR}(r_{p,i-1},e_{i-1})]^{3/2}}{M^{1/2} {[\delta e_{i}^{\GR}(r_{p,i-1},e_{i-1})]^{3/2}}}.
\end{align}
Using the above functions, we can relate the desired properties of the GW burst following the $i^{th}$ orbit to
the parameters of the preceding orbit via:
\begin{align}
\label{f2-exact}
f_{i}^{\GR} &=  \frac{M^{1/2} \left[2 - \delta e_{i}^{\GR}(r_{p,i-1},e_{i-1})\right]^{1/2}}{2 \pi [r_{p,i}^{\GR}(r_{p,i-1},e_{i-1})]^{3/2}},
\\
\label{deltat2-exact}
\delta t_{i}^{\GR} &= \xi_{t}  \frac{[r_{p,i}^{\GR}(r_{p,i-1},e_{i-1})]^{3/2}}{M^{1/2} \left[2 - \delta e_{i}^{\GR}(r_{p,i-1},e_{i-1})\right]^{1/2}} ,
\\
\label{deltaf2-exact}
\delta f_{i}^{\GR} &= \xi_{f}  \frac{M^{1/2}\left[2 - \delta e_{i}^{\GR}(r_{p,i-1},e_{i-1})\right]^{1/2}}{2 \pi [r_{p,i}^{\GR}(r_{p,i-1},e_{i-1})]^{3/2}} ,
\end{align}
We have added here and in Eqs.~\eqref{rp2} to~\eqref{t2-exact} a GR superscript to remind us that these mapping relations are the predictions of Einstein's theory.  

In practice, the burst sequence given by iteration of these equations must be terminated at some point. 
For the derivation of the ppE deformations this is not necessary, though we briefly mention
some of the issues that might need to be considered. First, regardless of eccentricity, when the
pericenter separation approaches an effective inner-most stable orbit, the binary will merge. If the
eccentricity is still large at this moment, the plunge/merger/ringdown could be modeled as the final
burst, though it would have different characteristics (and ppE deformation) in the time-frequency plane 
than the fly-by bursts.
For orbits with large initial pericenter passage, the orbit can become sufficiently close to 
circular before merger that the binary begins to emit high luminosity GWs essentially 
continuously; i.e. in the burst model, the time-frequency tiles will begin to touch, or even overlap if $(\xi_t,\xi_f)$ are large enough. 
This is not {\em a-priori} a problem (see for example~\cite{Neil-Tyson}), though
alternatively it may be more efficient at this stage to analyze the rest of the inspiral with a traditional
matched filtering algorithm, using small eccentricity waveform templates. Such templates would need to be ppE-deformed
as well, and it would be necessary (and an interesting exercise on its own) 
to map the burst-ppE parameters to those of ppE eccentric inspiral templates.

\subsection{The Inverse Problem and Degeneracies}

The parameter vector that describes the burst algorithms is $\lambda^{a} = (t_{0,\ldots,N},f_{0,\ldots,N},\delta t_{0,\ldots,N}, \delta f_{0,\ldots,N})$ for $N$ bursts detected. That is, each burst is fully described by $4$ parameters, thus leading to $4N$ pieces of information. However, an inspiraling, eccentric orbit in GR to leading PN order (neglecting spins) is described by a much smaller vector $\lambda^{a}_{\rm sys} = (M,\eta, e_{0}, r_{p,0})$. Given $\lambda^{a}$, the questions then are the following: how do we map back to $\lambda^{a}_{\rm sys}$; and what are the minimum number of bursts needed to specify this inverse map uniquely, if it exists? Answers to practical versions of these questions would require a more thorough analysis where detector noise is considered; here we just give a simple counting argument, and caution about degeneracies if a simple model, such as the one we use here, is employed.

Detection of the first burst provides four pieces of information, $(t_{1},f_{1},\delta t_{1}, \delta f_{1})$. The parameter $t_{1}$ is unrelated to any physical parameters. The parameter $f_{1}$ (for example) depends on $M$, $r_{p,0}$ and $e_{0}$, so it can be used to infer one component of $\lambda^{a}_{\rm sys}$. The parameters $\delta      t_{1}$ and $\delta f_{1}$ both depend on $\tau_{\GW}$, which in turn depends only on $f_{1}$; thus they do not provide  independent information. If a second burst is detected, four  more pieces of information are obtained: $(t_{2}, f_{2}, \delta t_{2}, \delta f_{2})$. As before, measuring one of $(f_2,\delta t_{2}, \delta f_{2})$ uniquely determines the other two, but now $t_{2}$ does depend on the system parameters. Hence in principle we have two new independent pieces of information from the second burst. Similarly for a third burst, but we know there are at most four independent numbers in $\lambda^{a}_{\rm sys}$ here, and hence three consecutive bursts should suffice to determine it.

This counting argument assumes there are no degeneracies amongst the parameters, which is not the case if leading PN order expressions are used to construct the algorithm. In that case, there is a degeneracy between $M$, $\eta$ and $r_{p}$, and the burst algorithm actually only depends on the physical parameters $\bar{\lambda}^{a}_{\rm sys} = ({\cal{M}}, {\cal{R}}_{0}, e_{0})$, where ${\cal{M}} = M \eta^{3/5}$ is the chirp mass and ${\cal{R}} = r_{p}^{3/2}/M^{1/2}$ is the the leading PN order expression for the radius of curvature of the binary. In principle two bursts would allow complete determination of $\bar{\lambda}^{a}_{\rm sys}$. This reduced parameter dependence can be seen most clearly if we expand the algorithm in both $M/r_{p} \ll 1 \gg \delta e$ to find
\begin{align}
t_{i} &= t_{i-1} + \frac{2 \pi {\cal{R}}_{i-1}^{3/2}}{\delta e_{i-1}^{3/2}} \left[1 + {\cal{O}}\left(\frac{{\cal{M}}^{5/2}}{{\cal{R}}_{i-1}^{5/2}}, \delta e_{i-1}\right)\right]\,,
\\
f_{i} &= \frac{1}{\sqrt{2}\pi {\cal{R}}_{i-1}} \left[ 1 + {\cal{O}}\left(\frac{{\cal{M}}^{5/2}}{{\cal{R}}_{i-1}^{5/2}}, \delta e_{i-1}\right) \right]\,,
\\
\delta t_{i} &= \delta t_{i-1} \left[1 + {\cal{O}}\left(\frac{{\cal{M}}^{5/2}}{{\cal{R}}_{i-1}^{5/2}}, \delta e_{i-1}\right)\right]\,,
\\
\delta f_{i} &= \delta f_{i-1} \left[1 + {\cal{O}}\left(\frac{{\cal{M}}^{5/2}}{{\cal{R}}_{i-1}^{5/2}}, \delta e_{i-1}\right) \right]\,,
\end{align}
The situation is very reminiscent to degeneracies that arise in the stationary-phase approximation to the Fourier transform of the response function for GWs emitted during quasi-circular inspirals: the Fourier phase only depends on ${\cal{M}}$ to leading PN-order, and not on both $m_{1}$ and $m_{2}$ independently. Of course, if one were to develop a burst algorithm beyond leading PN order, these degeneracies would be broken. 

\section{Modeling beyond GR}
\label{sec:modeling-beyond-GR}

In this section, we use the simple GR burst algorithm developed in the previous section to construct a ppE generalization. We consider model-independent deformations of GR that lead to modifications to the conservative dynamics of the form
\begin{align}
\label{E-eq}
E &= E_{\GR} \left[1 + \alpha \left(\frac{M}{r_p}\right)^{\bar{a}}\right]\,,
\\
\label{L-eq}
L &= L_{\GR} \left[1 + \beta \left(\frac{M}{r_p}\right)^{\bar{b}}\right]\,,
\end{align}
and modifications to the dissipative dynamics of the form
\begin{align}
\label{Edot-eq}
\dot{E} &= \dot{E}_{\GR} \left[1 + \gamma \left(\frac{M}{r_p}\right)^{\bar{c}}\right]\,,
\\
\label{Ldot-eq}
\dot{L} &= \dot{L}_{\GR} \left[1 + \delta \left(\frac{M}{r_p}\right)^{\bar{d}}\right]\,,
\end{align}
where $(E_{\GR},L_{\GR})$ are the GR orbital energy and angular momentum, 
while $(\dot{E}_{\GR},\dot{L}_{\GR})$ are the GR orbital energy and angular momentum flux. 

The modifications introduced above are reasonable when one is searching for strong-field modifications to GR in the following sense. They must be proportional to a variable that is small in the weak-field to be consistent with existing observations, yet large in the strong-field to offer a chance to be detectable with GW observations. A natural choice is the characteristic velocity of the system during emission, which for an eccentric inspiral occurs at pericenter and is inversely proportional to the pericenter radius. In fact, as we show in Sec.~\ref{sec:burst-in-mod-gravity}, the above deformations capture the leading order effects from certain modified gravity theories.  Also, as discussed in Appendix~\ref{one-body}, an equivalent way to arrive at these changes is through
polynomial deformations to the effective-one-body Lagrangian.

The post-Einsteinian corrections are parameterized by the amplitude coefficients 
$\delta \lambda^{a} \equiv (\alpha,\beta,\gamma,\delta) \in \Re$
and the exponent coefficients $\delta \ell^{a} \equiv (\bar{a},\bar{b},\bar{c},\bar{d}) \in \mathbb{Q}$.
In principle, the vector $\delta \lambda^{a}$ need not be constant, and in fact, it may be a function
of other orbital parameters, such as the eccentricity or spin. Nonetheless, if we focus
only on small GR deformations, and on orbits that are highly-eccentric, then $\delta \lambda^{a}$ will be 
a constant dependent only on the coupling parameters of the theory; we will work here to first-order
in $\delta \lambda^{a}$. On the other hand, $\delta \ell^{a}$ can have any magnitude, except that
$(\bar{a},\bar{b}) > 0$ if the GR corrections are to lead to well-behaved observables in the weak-field. 
If this were not the case, then the spacetime that generates such corrections would not be asymptotically flat. 
The subset $(\bar{c},\bar{d})$ can be less than zero, as is the case in theories of gravity with dipolar radiation. 

One could parameterize GR deviations with a different 
functional basis, such as natural logarithms, but a power-law basis is perhaps more natural, unless 
screening is present. In the next subsection, we will investigate how these parameterized 
post-Einsteinian corrections modify our burst algorithm; most of this analysis will closely follow
the work presented in Sec.~\ref{sec:GR-modeling}.

\subsection{Size of Tiles}

The first ingredient of the algorithm is the size of the tiles, which is given by the characteristic GW time, $\tau_{\GW}^{\ppE}$, defined in Eq.~\eqref{tauGW-def}. In a modified gravity theory, however, the periastron velocity is not the same function of $(r_{p},e)$ as in GR, because of the modification to the binding energy in Eq.~\eqref{E-eq}. In Eq.~\eqref{vppE}, we present the corrected expression, which then leads to 
\begin{align}
\label{eq:ppEtau}
\tau_{\GW}^{\ppE} 
&= \frac{r_{p}^{3/2}}{\sqrt{M (1 + e)}} \left[1 - \frac{1}{3} \alpha \left(\frac{M}{r_{p}} \right)^{\bar{a}} - \frac{1}{3} \beta \left(\frac{M}{r_p}\right)^{\bar{b}}\right]\,.
\end{align}
The characteristic GW frequency, $f_{\GW}^{\ppE}$, then follows straightforwardly from Eq.~\eqref{fGW-def}.

The next step is to find the dimensions of the tiles. Using the same notation as in Sec.~\ref{sec:GR-modeling}, we easily find
\begin{align}
\label{eq:deltat-ppE}
\delta t^{\ppE} &= \delta t^{\GR} \left[1 - \frac{1}{3} \alpha \left(\frac{M}{r_{p}}\right)^{\bar{a}} - \frac{1}{3} \beta \left(\frac{M}{r_{p}}\right)^{\bar{b}}\right]\,,
\\
\label{eq:deltaf-ppE}
\delta f^{\ppE} &= \delta f^{\GR} \left[1 + \frac{1}{3} \alpha \left(\frac{M}{r_{p}}\right)^{\bar{a}} + \frac{1}{3} \beta \left(\frac{M}{r_{p}}\right)^{\bar{b}}\right]\,,
\end{align}
where $\delta t^{\GR}$ and $\delta f^{\GR}$ are given by Eqs.~\eqref{deltat} and~\eqref{deltaf} respectively. 

\subsection{Mapping Between Tiles}
We now seek to determine how the mapping between tiles is modified in our ppE model. Just as in GR, we need to determine how the orbital quantities map from one burst to the next: $(r_{p,i}, e_{i}) \rightarrow (r_{p,i+1}, e_{i+1})$. Let us begin by mapping the ppE deformations of the energy and angular momentum to the pericenter distance and eccentricity. Using the leading, Newtonian order expressions for $E_{\rm GR}$ and $L_{\rm GR}$ in a weak-field expansion, we find
\begin{align}
\label{rp-ppE}
r_{p}^{\rm ppE} &= \frac{M}{2 {\cal{E}}} \left(1 - \sqrt{1 - 2 {\cal{E}} {\cal{L}}^{2}}\right) 
\nonumber \\
& \times \left[1 - \frac{1 - {\cal{E}} {\cal{L}}^{2} - \sqrt{1 - 2 {\cal{E}} {\cal{L}}^{2}}}{(1 - \sqrt{1 - 2 {\cal{E}} {\cal{L}}^{2}}) \sqrt{1 - 2 {\cal{E}} {\cal{L}}^{2}}} \alpha \left(\frac{M}{r_{p}}\right)^{\bar{a}}
\right.
\nonumber \\
& \left.
- \frac{2 {\cal{E}} {\cal{L}}^{2}}{(1 - \sqrt{1 - 2 {\cal{E}} {\cal{L}}^{2}}) \sqrt{1 - 2 {\cal{E}} {\cal{L}}^{2}}} \beta \left(\frac{M}{r_{p}}\right)^{\bar{b}} 
\right.
\nonumber \\
& \left.
+ {\cal{O}}(\delta\lambda^{a}\delta\lambda_{a}) \right.\Bigg]\,,
\\
\label{deltae-ppE}
e &= \sqrt{1 - 2 {\cal{E}} {\cal{L}}^{2}} \left\{1 + \frac{{\cal{E}} {\cal{L}}^{2}}{1 - 2 {\cal{E}} {\cal{L}}^{2}}
\right.
\nonumber \\
& \left.
\times \left[\alpha \left(\frac{M}{r_{p}}\right)^{\bar{a}} + 2 \beta \left(\frac{M}{r_{p}}\right)^{\bar{b}} + 
{\cal{O}}(\delta\lambda^{a}\delta\lambda_{a})\right] \right\}\,,
\end{align}
The pericenter distance and the eccentricity of the $i^{th}$ burst is thus controlled by the dimensionless energy ${\cal{E}}$ and angular momentum ${\cal{L}}$ of that burst, just as in Eqs.~\eqref{r-inv} and~\eqref{e-inv}, but with ppE modifications.

The next step is to find how the energy and angular momentum evolve from one burst to the next. We prescribe this evolution through the perturbative algorithm described below Eq.~\eqref{DeltaL}. The ppE corrected energy and angular momentum flux are given in Eqs.~\eqref{Edot-eq} and~\eqref{Ldot-eq} respectively, where we take $\dot{E}_{\rm GR}$ and $\dot{L}_{\rm GR}$ to leading, Newtonian order. We can then insert these expressions in Eqs.~\eqref{rp-ppE} and~\eqref{deltae-ppE} and linearize in $\delta \lambda^{a}$ and $\delta e_{i-1}$ to find
\begin{widetext}
\begin{align}
\label{rp2-ppE}
r_{p,i}^{\rm ppE}&(r_{p,i-1},e_{i-1}) = \nn\\
&r_{p,i-1} \left\{1 - \frac{59 \sqrt{2} \pi}{24} \eta \left(\frac{M}{r_{p,i-1}}\right)^{5/2} \left[1 + \frac{26}{59} \alpha \left(\frac{M}{r_{p,i-1}}\right)^{\bar{a}} - \frac{144}{59} \beta \left(\frac{M}{r_{p,i-1}}\right)^{\bar{b}} - \frac{85}{59} \gamma \left(\frac{M}{r_{p,i-1}}\right)^{\bar{c}} + \frac{144}{59} \delta \left(\frac{M}{r_{p,i-1}}\right)^{\bar{d}}\right]
\right.
\nonumber \\
& \left.
- \frac{121 \sqrt{2} \pi}{96} \eta \left(\frac{M}{r_{p,i-1}}\right)^{5/2} \delta e_{i-1} \left[1 + \frac{1142}{605} \alpha \left(\frac{M}{r_{p,i-1}}\right)^{\bar{a}} - \frac{2352}{605} \beta \left(\frac{M}{r_{p,i-1}}\right)^{\bar{b}} - \frac{1747}{605} \gamma \left(\frac{M}{r_{p,i-1}}\right)^{\bar{c}} + \frac{2352}{605} \delta \left(\frac{M}{r_{p,i-1}}\right)^{\bar{d}}\right]
\right. \nn \\ & \left.
+ {\cal{O}}\left[\left(\frac{M}{r_{p,i-1}}\right)^{\delta \ell^{a} + 7/2},\delta e_{i-1}^{2},\delta\lambda^{a}\delta\lambda_{a}\right]\right\}\,,
\\
\label{e2-ppE}
\Delta \delta &e_{i,i-1}^{\rm ppE}(r_{p,i-1},e_{i-1}) = \frac{85 \sqrt{2} \pi}{12} \eta \left(\frac{M}{r_{p,i-1}}\right)^{5/2} \left[1 - 2 \alpha \left(\frac{M}{r_{p,i-1}}\right)^{\bar{a}} + \gamma \left(\frac{M}{r_{p,i-1}}\right)^{\bar{c}}\right] - \frac{181 \sqrt{2} \pi}{80} \eta \left(\frac{M}{r_{p,i-1}}\right)^{5/2} 
\nonumber \\
&  
\times \left[1 + \frac{118}{181} \alpha \left(\frac{M}{r_{p,i-1}}\right)^{\bar{a}} - \frac{480}{181} \beta \left(\frac{M}{r_{p,i-1}}\right)^{\bar{b}} - \frac{299}{181} \gamma \left(\frac{M}{r_{p,i-1}}\right)^{\bar{c}} + \frac{480}{181} \delta \left(\frac{M}{r_{p,i-1}}\right)^{\bar{d}}\right]\delta e_{i-1} 
\nn\\
&+ {\cal{O}}\left[\left(\frac{M}{r_{p,i-1}}\right)^{\delta \ell^{a} + 7/2},\delta e_{i-1}^{2},\delta\lambda^{a}\delta\lambda_{a}\right]\,,
\end{align}
\end{widetext}
where we have defined $\Delta \delta e_{i,i-1}^{\rm ppE} = \delta e_{i}^{\rm ppE} - \delta e_{i-1}$.

As before, we can now construct the mapping between tiles. By combining the above results, we find
\begin{align}
\label{eq:ppEexact1}
t_{i}^{\rm ppE} &= t_{i-1} + \Delta t_{i,i-1}^{\rm GR}(r_{p,i}, e_{i}) \left[1 - \alpha \left(\frac{M}{r_{p,i}}\right)^{\bar{a}} 
\right. \nn \\ &\left. 
+ {\cal{O}}\left(\delta\lambda_{a} \delta \lambda^{a}\right) \right.\Bigg]\,,
\\
f_{i}^{\rm ppE} &= f_{i}^{\rm GR}(r_{p,i}, e_{i}) 
\left[1 + \frac{1}{3} \delta {\cal{F}}_{} + {\cal{O}}\left(\delta\lambda_{a} \delta \lambda^{a}\right) \right]\,,
\\
\delta t_{i}^{\rm ppE} &= \delta t_{i}^{\rm GR}(r_{p,i}, e_{i}) 
\left[1 - \frac{1}{3} \delta {\cal{F}}_{}  + {\cal{O}}\left(\delta\lambda_{a} \delta \lambda^{a}\right) \right]\,,
\\
\label{eq:ppEexact4}
\delta f_{i}^{\rm ppE} &= \delta f_{i}^{\rm GR}(r_{p,i}, e_{i}) 
\left[1 + \frac{1}{3} \delta {\cal{F}}_{}  + {\cal{O}}\left(\delta\lambda_{a} \delta \lambda^{a}\right) \right]\,,
\end{align}
where we have defined
\be
\delta {\cal{F}}  \equiv  \alpha \left(\frac{M}{r_{p,i}}\right)^{\bar{a}} + \beta \left(\frac{M}{r_{p,i}}\right)^{\bar{b}}\,.
\ee
and where $\Delta t_{i,i-1}^{\GR}$, $f_{i}^{\GR}$, $\delta t_{i}^{\GR}$ and $f_{i}^{\GR}$ are the GR mappings of Eqs.~\eqref{t2-exact}--\eqref{deltaf2-exact}. Recall that these quantities are all implicit functions of $(r_{p,i-1},e_{i-1})$, with the mapping $(r_{p,i}, e_{i}) \rightarrow (r_{p,i-1}, e_{i-1})$ given in Eqs.~\eqref{rp2-ppE} and~\eqref{e2-ppE}. The time to the next burst is obtained by using the ppE-modified orbital period in Eq.~\eqref{eq:ppEexact1}. The frequency to the next burst is given by the inverse of the ppE-modified GW characteristic time in Eq.~\eqref{eq:ppEtau}, while the size of the burst windows in time-frequency space is given in Eqs.~\eqref{eq:deltat-ppE} and~\eqref{eq:deltaf-ppE}. 

\subsection{A parameterized-post Einsteinian Burst Framework}
\label{subsect:ppE-burst}

The previous subsection inspires the ppE burst algorithm presented in Eqs.~\eqref{eq:ppEexact1-new}-\eqref{eq:deltas-new}, where we have redefined the ppE amplitude and exponent coefficients.  Notice that we only need two sets of parameters in Eqs.~\eqref{eq:ppEexact1-new}-\eqref{eq:ppEexact4-new}, $(\alpha_{\ppE},\bar{a}_{\ppE})$ and $(\beta_{\ppE},\bar{b}_{\ppE})$, since the frequency of the next burst, the temporal and the frequency sizes of the burst window are all controlled by a single quantity, the characteristic time $\tau_{\GW}$. 

The amplitude parameters  $(\alpha_{\ppE},\beta_{\ppE},\gamma_{\ppE},\delta_{\ppE})$ are, in principle, not only functions of the coupling constants of the theory, but also functions of the eccentricity of the $i^{th}$ orbit. When working in the high-eccentricity limit, one can expand these via
\be
\epsilon_{\ppE}(e_{i}) = \bar{\epsilon}_{\ppE,0} +  \bar{\epsilon}_{\ppE,1} \delta e_{i} + {\cal{O}}\left(\delta e_{i}^{2}\right),
\ee
where $\epsilon_{\ppE}(e_{i})$ is any of $(\alpha_{\ppE},\beta_{\ppE},\gamma_{\ppE},\delta_{\ppE})$, while 
$(\bar{\epsilon}_{\ppE,0},\bar{\epsilon}_{\ppE,1})$
are related constants.

One may wonder how the ppE parameters $(\alpha_{\ppE},\beta_{\ppE},\gamma_{\ppE},\delta_{\ppE})$ and $(\bar{a}_{\ppE},\bar{b}_{\ppE},\bar{c}_{\ppE},\bar{d}_{\ppE})$ in Eqs.~\eqref{eq:ppEexact1-new}-\eqref{eq:deltas-new} map to the deformations to the conservative and dissipative dynamics in Eqs.~\eqref{E-eq}-\eqref{Ldot-eq}. By comparing the former to the equations in the previous section, one finds that $\alpha_{\ppE} = -\alpha$ and $\bar{a}_{\ppE} = \bar{a}$. The mapping between the other parameters, however, depends on which sector is most dominant in PN theory, i.e.~which coefficient $(\bar{a},\bar{b},\bar{c},\bar{d})$ is largest. If one knows what the deformations in Eqs.~\eqref{E-eq}-\eqref{Ldot-eq} are, one can then construct a one-to-one mapping between these and the ppE coefficients:
\begin{align}
\alpha_{\ppE} &= - \alpha\,, \qquad \bar{a}_{\ppE} = \bar{a}\,,
\\
\beta_{\ppE} \left(\frac{M}{r_{p}}\right)^{\bar{b}_{\ppE}} &= {\rm{LO}} \left[\frac{\alpha}{3} \left(\frac{M}{r_{p}}\right)^{\bar{a}} + \frac{\beta}{3} \left(\frac{M}{r_{p}}\right)^{\bar{b}}\right] \,,
\\
\gamma_{\ppE} \left(\frac{M}{r_{p}}\right)^{\bar{c}_{\ppE}} & = - \frac{59 \sqrt{2} \pi}{24} \eta \; {\rm{LO}} \left[\frac{26}{59} \alpha \left(\frac{M}{r_{p}}\right)^{\bar{a}+5/2} 
\right.
\nn \\
&\left. 
- \frac{144}{59} \beta \left(\frac{M}{r_{p}}\right)^{\bar{b}+5/2} 
- \frac{85}{59} \gamma \left(\frac{M}{r_{p}}\right)^{\bar{c}+5/2} 
\right.
\nn \\
&\left. 
+ \frac{144}{59} \delta \left(\frac{M}{r_{p}}\right)^{\bar{d}+5/2}\right]\,,
\\
\delta_{\ppE} \left(\frac{M}{r_{p}}\right)^{\bar{d}_{\ppE}} & = {\rm{LO}} \left[- 2 \alpha \left(\frac{M}{r_{p}}\right)^{\bar{a}} + \gamma \left(\frac{M}{r_{p}}\right)^{\bar{c}}\right]\,,  
\end{align}
where $\rm{LO}$ stands for the leading-order term in an $M/r_{p} \ll 1$ expansion, which of course depends on the value of $(\bar{a},\bar{b},\bar{c},\bar{d})$. 

With this algorithm at hand, one can formulate the following tests of GR. Given a detection of N bursts, one obtains the observables $(t_{0,\ldots,N},f_{0,\ldots,N},t_{0,\ldots,N},f_{0,\ldots,N})$. The data should be completely described by the system parameters $\bar{\lambda}^{a}_{\rm sys} = ({\cal{M}},{\cal{R}}_{0},e_{0})$, the ppE amplitude parameters $(\alpha_{\ppE},\beta_{\ppE},\gamma_{\ppE},\delta_{\ppE})$ and the ppE exponent parameters $(\bar{a}_{\ppE},\bar{b}_{\ppE},\bar{c}_{\ppE},\bar{d}_{\ppE})$. Provided we detect enough bursts (i.e.~$N$ is large enough), one should be able to study whether the data prefers a GR model, i.e.~one with $(\alpha_{\ppE},\beta_{\ppE},\gamma_{\ppE},\delta_{\ppE}) = (0,0,0,0)$, or not. As in the quasi-circular case, one expects there to be degeneracies between the ppE parameters~\cite{cornishsampson}. The extent to which such degeneracies affect the tests of GR proposed here will be investigated in detail elsewhere. One could of course also extend the ppE parameterization to include more ppE coefficients, but this may dilute the strength of the GR tests. 

\section{Burst Models in Modified Gravity}
\label{sec:burst-in-mod-gravity}

In this section, we study GW bursts in two specific theories of gravity: EDGB and BD theory. 
In particular, this will require the calculation of the binding energy, angular moment, energy flux
and angular momentum flux for eccentric inspirals, some of which had not been calculated before.
Such a study will allow us to determine the efficiency of the ppE modifications to capture 
specific GR modifications.

\subsection{Einstein-Dilaton-Gauss-Bonnet Gravity}

EDGB is a quadratic modified gravity theory that corrects the Einstein-Hilbert action through the product of a dynamical scalar field $\vartheta$ and the Gauss-Bonnet invariant. The field equations for the metric and the scalar field are given for example in~\cite{yunesstein}. Black hole and neutron star solutions were found in~\cite{yunesstein,Pani:2011xm}, while binary systems were analyzed in~\cite{yunesstein,quadratic}. This theory contains one coupling constant $\xi$, with dimensions of length to the fourth power; the theory reduces to GR in the limit $\xi \to 0$. The strongest constraint on $\xi$ has been derived from observations of low-mass X-ray binaries in~\cite{kent-LMXB}, namely $\xi^{1/4} \lesssim 5 \times 10^{3} \; {\rm{m}}$. One of the most important EDGB modifications to GR occurs when considering binary black holes, which emit gravitational dipole radiation due to the excitation of the scalar field, which induces $-1$PN order corrections~\cite{quadratic}. For neutron star  binaries, it is not clear that gravitational dipole radiation is excited, so in the rest of this section, we will consider only bursts emitted by systems where at least one of the binary components is a black hole. 

Let us first consider the effective-one-body problem for a particle in an eccentric orbit about a compact object, with \emph{specific} orbital energy $E$ and magnitude of (the z-component of) angular momentum $L$. Recall that when mapping the two-body problem to an effective-one-body problem, the mass of the particle is simply the reduced mass of the black hole binary $\mu=m_{1} m_{2}/(m_{1}+m_{2})$, while the mass of the compact object is the total mass $M = m_{1}+m_{2}$. For a black hole binary, the effective potential in EDBG is given by Eq. (13) in~\cite{Yunes:2011we}: $V_{\rm eff} = V_{\rm eff}^{\rm GR} + \delta V_{\rm eff}$ where $V_{\rm eff}^{\rm GR}$ is the effective potential in GR~\cite{Carrol}, while $\delta V_{\rm eff} = -E^{2} h/2 - V_{\rm eff}^{\rm GR} k/2$, where $f = 1 - 2M/r$ is the so-called Schwarzschild factor, with~\cite{Yunes:2011we}
\begin{align}
h &= \frac{\zeta}{3f} \left(\frac{M}{r}\right)^{3} \left[1 + 26 \frac{M}{r} + {\cal{O}}\left(\frac{M^{2}}{r^{2}}\right)\right]\,,
\\
k &= - \frac{\zeta}{f} \left(\frac{M}{r}\right)^{2} \left[1 + \frac{M}{r} + {\cal{O}}\left(\frac{M^{2}}{r^{2}}\right)\right]\,,
\end{align}
and $\zeta \equiv \xi/M^{4}$. 

The orbital energy and angular momentum for an eccentric orbit can be found by determining the turning points in the orbits (see also Appendix~\ref{one-body}). The EDGB corrections for a black hole binary are then 
\begin{align}
\frac{\delta E}{E_{\GR}} &= - \frac{\zeta}{12} \frac{(1-e)^{2}}{1+e} \left(\frac{M}{r_{p}}\right)^{3} 
+ {\cal{O}}\left(\frac{M^{4}}{r_{p}^{4}}\right)\,,
\nn \\
&= -\frac{\zeta}{24} \delta e^{2} \left(\frac{M}{r_{p}}\right)^{3} \left[1 + \frac{1}{2} \delta e\right] 
+ {\cal{O}}\left(\frac{M^{4}}{r_{p}^{4}},\delta e^{4}\right)\,,
\\ 
\frac{\delta L}{L_{\GR}} &= - \frac{\zeta}{12} \frac{3 + e^{2}}{(1+e)^{2}} \left(\frac{M}{r_{p}}\right)^{2}
+ {\cal{O}}\left(\frac{M^{3}}{r_{p}^{3}}\right)\,,
\nn \\
& = - \frac{\zeta}{12} \left(\frac{M}{r_{p}}\right)^{2} \left[1 + \frac{1}{2} \delta e\right]
+ {\cal{O}}\left(\frac{M^{3}}{r_{p}^{3}},\delta e^{2}\right)\,,
\end{align} 
where $E_{\GR}$ and $L_{\GR}$ can be found e.g. in~\cite{Barack:2011ed}, and where we have expanded to leading-order in $\zeta$ and $M/r_{p}$. In the second lines of the above equations, we have also expanded in $\delta e \ll 1$. We have checked that these results agree exactly with those in~\cite{Yunes:2011we} when one takes the circular limit $e \to 0$.

Let us now consider how the orbital energy and angular momentum change due to radiation losses. When considering binaries where at least one of the components is a black hole, the dominant radiation loss is due to scalar-field emission [see e.g.~Eqs.~(122) in~\cite{quadratic} and (B23) in~\cite{Yagi:2013mbt}]:
\begin{align}
\dot{E}^{(\vartheta)} &= - \frac{4 \pi}{3} \beta \left< \ddot{D}_{i} \ddot{D}^{i} \right>\,,
\qquad
\dot{L}_{i}^{(\vartheta)} = \frac{4 \pi}{3} \beta \epsilon_{ijk} \left< \ddot{D}_{j} \dot{D}_{k} \right>\,,
\end{align}
where $D^{i} = q_{1} x_{1}^{i} + q_{2} x_{2}^{i}$ is an induced dipole moment, with $q_{1,2} =  m_{1,2} [\zeta_{1,2}/(4 \pi \beta)]^{1/2}$ the scalar charges of the compact objects, $\zeta_{1,2} = \xi/m_{1,2}^{4}$, $x_{1,2}^{i}$ the position vector relative to the center of mass, and $\left<\right>$ stands for orbit averaging. The above expressions have already been angle averaged (i.e. integrated over solid angle at spatial infinity). Carrying out the orbit average, we find
\begin{align}
\frac{\dot{E}^{(\vartheta)}}{\dot{E}_{\rm GR}} &= \frac{5}{96} {\cal{S}}_{\rm EDGB}^{2} \frac{(1+e)(1+\frac{1}{2}e^{2})}{1 + \frac{73}{24}e^{2} + \frac{37}{96}e^{4}}\left(\frac{M}{r_{p}}\right)^{-1} + {\cal{O}}(1)\,,
\nn \\
&= \frac{3}{85} {\cal{S}}_{\rm EDGB}^{2} \left(\frac{M}{r_{p}}\right)^{-1} \left( 1 + \frac{1417}{2550} \delta e\right)
+ {\cal{O}}(1,\delta e^{2})\,,
\\
\frac{\dot{L}^{(\vartheta)}}{\dot{L}_{\rm GR}} &= \frac{5}{96}{\cal{S}}_{\rm EDGB}^{2} \frac{1+e}{1+\frac{7}{8}e^{2}} \left(\frac{M}{r_{p}}\right)^{-1} + {\cal{O}}(1)\,,
\nn \\
&= \frac{1}{18} {\cal{S}}_{\rm EDGB}^{2} \left(\frac{M}{r_{p}}\right)^{-1} \left(1 + \frac{13}{30} \delta e\right)
+ {\cal{O}}(1,\delta e^{2})\,,
\end{align}
where $\dot{E}_{\GR}$ and $\dot{L}_{\GR}$ are the energy and angular momentum flux in GR~\cite{Peters:1964zz,PetersMathews,Carrol}, and where we have defined ${\cal{S}}_{\rm EDGB} = \zeta_{1}^{1/2} - \zeta_{2}^{1/2}$.  Again, in the second lines of the above equations we have expanded in $\delta e \ll 1$. Notice that when $m_{1} = m_{2}$, then ${\cal{S}}_{\rm EDGB} = 0$ and dipole radiation vanishes. We have checked that these results agree exactly with those of~\cite{Yagi:2013mbt} in the circular limit $e \to 0$. For mixed neutron star-black hole binaries, we set $\zeta_{1} = \zeta_{\rm BH} = \xi/m_{\rm BH}^{4}$ and $\zeta_{2} = \zeta_{\rm NS} = 0$.

We can then easily map between the EDGB deformations described above and the ppE deformations of Eqs.~\eqref{E-eq}-\eqref{Ldot-eq}:
\begin{align}
\delta \lambda^{a}_{\rm EDGB} &= \left[-\frac{\zeta}{12} \frac{(1-e)^{2}}{1+e}, -\frac{\zeta}{12} \frac{3+e^{2}}{(1+e)^{2}},  
\right.
\nonumber \\
& \left.
\frac{5}{96} {\cal{S}}_{\rm EDGB}^{2} \frac{(1+e)(1+\frac{1}{2}e^{2})}{1+\frac{73}{24}e^{2}+\frac{37}{96}e^{4}},
\right.
\nonumber \\
& \left.
\frac{5}{96}  {\cal{S}}_{\rm EDGB}^{2}  \frac{1 + e}{1 + \frac{7}{8}e^{2}}\right]\,,
\end{align}
and $\delta \ell^{a}_{\rm EDGB} = \left(3, 2, -1, -1\right)$. Notice that $\delta \lambda^{a}_{\rm EDGB}$ are all functions not only of the coupling constant of the theory $\zeta$, but also of the eccentricity. Expanding to leading-order in $\delta e \ll 1$, we find 
\begin{align}
\label{lambda-EDGB}
\delta \lambda^{a}_{\rm EDGB} &= \left[0,-\frac{\zeta}{12},\frac{3}{85} S_{\rm EDGB}^{2}, \frac{1}{18} S_{\rm EDGB}^{2}\right]\,,
\end{align}
with remainders of ${\cal{O}}(\delta e)$. 

When considering GW bursts emitted by binaries where at least one of the components is a black hole, we then have all the necessary ingredients to construct the burst algorithm within EDGB. Combining our results with Eqs.~\eqref{rp2-ppE} and \eqref{e2-ppE} and working to leading, PN order and to first order in $\delta e$, we find
\begin{align}
t_{i}^{\rm EDGB} &= t_{i-1} + \Delta t_{i,i-1}^{\rm GR}(r_{p,i},\delta e_{i}) \left[1 + {\cal{O}}\left(\zeta^{2}, \delta e_{i}^{2}\right)\right]\,,
\\
f_{i}^{\rm EDGB} &= f_{i}^{\rm GR}(r_{p,i},\delta e_{i}) \left[1 - \frac{\zeta}{36} \left(\frac{M}{r_{p,i-1}}\right)^{2} 
\right.
\nn \\
& \left. + {\cal{O}}\left(\zeta^{2}, \delta e_{i}\right)\right]\,,
\\
\delta t_{i}^{\rm EDGB} &= \delta t_{i}^{\rm GR}(r_{p,i},\delta e_{i}) \left[1 + \frac{\zeta}{36} \left(\frac{M}{r_{p,i-1}}\right)^{2} 
\right.
\nn \\
& \left. + {\cal{O}}\left(\zeta^{2}, \delta e_{i}\right)\right]\,,
\\
\delta f_{i}^{\rm EDGB} &= \delta f_{i}^{\rm GR}(r_{p,i},\delta e_{i}) \left[1 - \frac{\zeta}{36} \left(\frac{M}{r_{p,i-1}}\right)^{2} 
\right.
\nn \\
& \left. + {\cal{O}}\left(\zeta^{2}, \delta e_{i}\right)\right]\,.
\end{align}
Recall that $\Delta t_{i,i-1}^{\rm GR}$, $f_{i}^{\rm GR}$, $\delta t_{i}^{\rm GR}$, $\delta f_{i}^{\rm GR}$ are functions of $r_{p,i}$ and $\delta e_{i}$, which are related to the parameters of the previous burst via
\begin{align}
\frac{r_{p,i}^{\rm EDGB} \left(r_{p,i-1}, \delta e_{i-1}\right)}{r_{p,i}^{\rm GR} \left(r_{p,i-1}, \delta e_{i-1}\right) } &= 1 - \frac{5 \sqrt{2} \pi}{24} \eta {\cal{S}}_{\rm EDGB}^{2} \left(\frac{M}{r_{p,i-1}}\right)^{3/2} 
\nn \\
&+ {\cal{O}}\left(\zeta^{2}, \delta e_{i-1}\right)
\\
\frac{\Delta \delta e_{i,i-1}^{\rm EDGB} \left(r_{p,i-1}, \delta e_{i-1}\right)}{\Delta \delta e_{i,i-1}^{\rm GR} \left(r_{p,i-1}, \delta e_{i-1}\right)} &= 1 + \frac{3}{85} {\cal{S}}_{\rm EDGB}^{2} \left(\frac{M}{r_{p,i-1}}\right)^{-1} 
\nn \\
&+ {\cal{O}}\left(\zeta^{2}, \delta e_{i-1}\right)
\end{align}
where recall that $\Delta \delta e_{i,i-1}^{\rm EDGB} =  \delta e_{i}^{\rm EDGB} - \delta e_{i-1}$.

Notice that the above EDGB modified burst algorithm maps exactly to the ppE burst model of Sec.~\ref{subsect:ppE-burst}. First, focusing on the temporal mapping of the burst tiles, we see that
\begin{align}
\alpha_{\ppE} = 0\,,
\qquad
\bar{a}_{\ppE} = 3\,. 
\end{align}
Next, notice that $\bar{a} = 3 > \bar{b} = 2$ in EDGB, and thus for the frequency mapping
\begin{align}
\beta_{\ppE} = -\frac{\zeta}{36}\,,
\qquad
\bar{b}_{\ppE} = 2\,. 
\end{align}
For the eccentricity mapping, we find that 
\begin{align}
\delta_{\ppE} = \frac{3}{85} {\cal{S}}_{\rm EDGB}^{2}\,,
\qquad
\bar{d}_{\ppE} = -1\,, 
\end{align}
since $\bar{c} = -1 < \bar{a}$ in EDGB. With this, it then becomes clear that 
\begin{align}
\gamma_{\ppE} = -\frac{5 \sqrt{2} \pi}{24} \eta {\cal{S}}_{\rm EDGB}^{2}\,,
\qquad
\bar{c}_{\ppE} = 3/2\,, 
\end{align}
because $5/2 + \bar{c} = 3/2$ in EDGB. We thus see that the EDGB burst model can be mapped to the ppE model to leading-order in the EDGB deformation parameter. 

As mentioned at the beginning of this subsection, one must be careful when apply the above ppE corrections to systems that contain neutron stars, since dipolar emission may be suppressed. Thus, in a mixed black hole-neutron star system, only the black hole would emit dipole radiation. Although the scalar charges for neutron stars have not yet been computed explicitly, it is expected that they will introduce corrections that are higher PN order compared to the dipolar radiation corrections described above when in presence of a black hole. This is why, when considering mixed binary systems above, we set $\zeta_{2} = \zeta_{\rm NS} = 0$ and only consider the effect of the black hole.

\subsection{Brans-Dicke Theory of Gravity}

BD theory~\cite{Brans:1961sx} is a particular scalar-tensor theory of gravity that has been extensively studied in the past (for a review, see e.g.~\cite{TEGP}). In this theory, the Ricci scalar in the Einstein-Hilbert action is multiplied by a dynamical scalar field $\phi$, which has the effect of effectively promoting Newton's gravitational constant $G$ to a function of spacetime. The field equations for the metric and the scalar field can be found e.g.~in~\cite{Brans:1961sx,TEGP,Will:1989sk}. Neutron stars in BD theory have been found in~\cite{Will:1989sk,Novak:1997hw}; black holes are not modified from their GR solution~\cite{Hawking:1972qk,Sotiriou:2011dz,Faraoni:2013iea}. Binary systems in BD theory were studied e.g.~in~\cite{Will:1989sk,Yunes:2011aa}. This theory possesses a dimensionless coupling constant $\omega_{\rm BD}$; the theory reduces to GR in the limit $\omega_{\rm BD} \to \infty$. The strongest constraint on the BD coupling constant comes from tracking of the Cassini spacecraft, which requires $\omega_{\rm BD} > 4 \times 10^{4}$~\cite{Bertotti:2003rm}. One of the most important BD modifications to GR is that the dynamical BD field induces dipolar emission of energy and angular momentum, just as in EDGB. Unlike EDGB, however, the conservative dynamics are not modified to leading PN order. 

Let us then consider the rate of change of the binding energy and angular momentum. Since the relation between the orbital period and the binding energy in BD is unchanged from GR, we can easily calculate the luminosity through $\dot{T}_{\rm orb}/T_{\rm orb} = -\frac{3}{2} \dot{E}/E$. Using the leading PN order expression for $\dot{T}_{\rm orb}$ (the second term in Eq. (2.26) of~\cite{Will:1989sk}), we can then find
\begin{align}
\frac{\dot{E}^{(\phi)}}{\dot{E}_{\GR}} &= \frac{5}{48} \frac{{\cal{S}}_{\rm BD}^{2}}{\omega_{\rm BD}} \left(\frac{M}{r_{p}}\right)^{-1} \frac{(1+e)(1+\frac{1}{2}e^{2})}{1+\frac{73}{24}e^{2}+\frac{37}{96}e^{4}}
\\
&= \frac{6}{85} \frac{{\cal{S}}_{\rm BD}^{2}}{\omega_{\rm BD}} \left(\frac{M}{r_{p}}\right)^{-1} \left(1 + \frac{1417}{2550} \delta e\right) + {\cal{O}}\left(1,\delta e^{2}\right)
\end{align}
where we have linearized in $1/\omega_{\rm BD}$, and we have defined ${\cal{S}}_{\rm BD} = s_{2} - s_{1}$, with $s_{1,2}$ the star's sensitivities. Alternatively, one can obtain the above by orbit averaging Eq. (6.16) in~\cite{Mirshekari:2013vb}, as we have checked explicitly. Observe that the modification to GR is proportional to the difference of the sensitivities squared, which vanishes for black hole binaries in which $s_{1} = 0.5 = s_{2}$. Similarly, for neutron star binaries, $s_{1} \approx s_{2}$ and the BD modification will be suppressed. For this reason, the largest BD modifications will be produced by black hole-neutron star binaries.  

Let us now focus on the rate of change of orbital angular momentum, a quantity that had not yet been calculated in the literature. To leading PN order and again linearizing in $1/\omega_{\rm BD}$,  this quantity is controlled by the BD field via
\begin{equation}
\dot{{L}}^{(\phi)}_{z} = \frac{\omega_{\rm BD}}{8 \pi} \lim_{R \rightarrow \infty} \int_{S^{2}_{R}} \left<\dot{\phi} \; (\vec{n} \times \vec{\nabla})_{z} \phi \right> R^{3} d\Omega\,,
\end{equation}
where $\vec{n}$ and $R$ are the unit vector and distance from the center of mass to a field point respectively, $\vec{\nabla}$ and $\times$ are the flat-space, spatial gradient and cross-product operators respectively (the latter of which is to be evaluated in the $\hat{z}$-direction, normal to the orbital plane) and the cosmological value of the scalar field is $\phi_{0} = 1 + {\cal{O}}(1/\omega_{\rm BD})$. Note that the above expression is both an angle average (the $d\Omega$ integral) and an orbit average (the $\left< \right>$ operator). The far-zone solution for the scalar field is~\cite{Will:1989sk}
\begin{equation}
\phi = -  2 \frac{{\cal{S}}_{\rm BD}}{\omega_{\rm BD}}  \frac{\eta M}{R} \left(\vec{n} \cdot \vec{v}\right)\,,
\end{equation}
to lowest order in the relative orbital velocity $\vec{v} = \dot{\vec{x}}_{2} - \dot{\vec{x}}_{1}$, where recall that $\eta$ is the binary's symmetric mass ratio, $M$ is the total mass and $\cdot$ is the flat-space inner product operator. Combining all of this and upon angle averaging, we obtain
\begin{align}
\dot{L}^{(\phi)}_{z} = \left<-\frac{2}{3} \frac{{\cal{S}}^{2}_{\rm BD}}{\omega_{\rm BD}} \frac{\eta M^{2}}{r^{3}} L \right>\,,
\end{align}
where $L$ is the orbital angular momentum and $r = |\vec{x}_{2} - \vec{x}_{1}|$ is the orbital separation. We now orbit average to obtain
\begin{align}
\frac{\dot{L}^{(\phi)}_{z}}{\dot{L}_{\GR}} &= \frac{5}{48} \frac{{\cal{S}}_{\rm BD}^{2}}{\omega_{\rm BD}} \left(\frac{M}{r_{p}}\right)^{-1} \frac{1+e}{1+\frac{7}{8}e^{2}}
\\
&= \frac{1}{9} \frac{{\cal{S}}_{\rm BD}^{2}}{\omega_{\rm BD}} \left(\frac{M}{r_{p}}\right)^{-1} \left(1 + \frac{13}{30} \delta e\right) + {\cal{O}}(1, \delta e^{2})
\end{align}
to linear order in $1/\omega_{\rm BD}$. To our knowledge, this result has not previously appeared in the literature. Furthermore, one can show that \emph{prior to orbit averaging}, $\dot{E}^{(\phi)}$ and $\dot{L}^{(\phi)}_{z}$ satisfy
\begin{equation}
\frac{\dot{E}^{(\phi)}}{\dot{L}^{(\phi)}_{z}} = \frac{a}{r} \frac{\Omega_{\rm orb}}{\sqrt{1-e^{2}}}
\end{equation}
In the above expression, $\dot{E}^{(\phi)}$ is given by Eq. (6.16) in~\cite{Mirshekari:2013vb} and $\Omega_{\rm orb}$ is the orbital frequency. As in the case of $\dot{E}^{(\phi)}$, note that $\dot{L}^{(\phi)}_{z}$ is suppressed when considering black hole or neutron star binaries, with the largest modifications arising from mixed binaries.

We can now  map the leading, PN order, BD modifications to the binding energy and angular momentum to the ppE deformations of  Eqs.~\eqref{E-eq}-\eqref{Ldot-eq}:
\begin{align}
\delta \lambda^{a}_{\rm BD} &= \left[0, 0, \frac{5}{48} \frac{{\cal{S}}_{\rm BD}^{2}}{\omega_{\rm BD}} \frac{(1+e)(1 + \frac{1}{2}e^{2})}{1 + \frac{73}{24} e^{2} + \frac{37}{96} e^{4}},
\right.
\nonumber \\
& \left.
\frac{5}{48} \frac{{\cal{S}}_{\rm BD}^{2}}{\omega_{\rm BD}} \frac{1+e}{1+\frac{7}{8}e^{2}} \right]\,,
\end{align}
and $\delta \ell^{a}_{\rm BD} = (0, 0, -1, -1)$. Once again, the dissipative sector corrections control the dominant ppE modifications. Notice also that $\delta \lambda^{a}_{\rm BD}$ clearly depends on the eccentricity, in addition to the BD coupling parameter. Expanding the above expressions to leading-order in the high-eccentricity limit, we find
\begin{align}
\label{lambda-BD}
\delta \lambda^{a}_{\rm BD} &= \left[0, 0, \frac{6}{85} \frac{{\cal{S}}_{\rm BD}^{2}}{\omega_{\rm BD}}, \frac{1}{9} \frac{{\cal{S}}_{\rm BD}^{2}}{\omega_{\rm BD}}\right]\,,
\end{align}
with remainders of ${\cal{O}}(\delta e)$. 

With the amplitude and exponent coefficients specified, we can now build the burst algorithm in BD theory. We find that the functionals of the burst algorithm are actually not modified from the GR result:  
\begin{align}
t_{i}^{\rm BD} &= t_{i} + \Delta t_{i,i-1}^{\rm GR}(r_{p,i}, \delta e_{i})
\\
f_{i}^{\rm BD} &= f_{i}^{\rm GR}(r_{p,i}, \delta e_{i})
\\
\delta t_{i}^{\rm BD} &= \delta t_{i}^{\rm GR}(r_{p,i}, \delta e_{i})
\\
\delta f_{i}^{\rm BD} &= \delta f_{i}^{\rm GR}(r_{p,i}, \delta e_{i})
\end{align}
but the mapping $(r_{p,i}, \delta e_{i}) \rightarrow (r_{p,i-1}, \delta e_{i-1})$ is modified via
\begin{align}
\frac{r_{p,i}^{\rm BD} \left(r_{p,i-1}, \delta e_{i-1}\right)}{r_{p,i}^{\rm GR} \left(r_{p,i-1}, \delta e_{i-1}\right)}  &= 1 - \frac{5 \sqrt{2} \pi}{12} \frac{{\cal{S}}_{\rm BD}^{2}}{\omega_{\rm BD}} \eta \left(\frac{M}{r_{p,i-1}}\right)^{3/2} 
\nn \\
&+ {\cal{O}}\left(\omega_{\rm BD}^{-2}, \delta e_{i-1}\right)\,,
\\
\frac{\Delta \delta e_{i,i-1}^{\rm BD} \left(r_{p,i-1}, \delta e_{i-1}\right)}{\Delta \delta e_{i,i-1}^{\rm GR} \left(r_{p,i-1}, \delta e_{i-1}\right)}  &= 1 + \frac{6}{85} \frac{{\cal{S}}_{\rm BD}^{2}}{\omega_{\rm BD}} \left(\frac{M}{r_{p,i-1}}\right)^{-1}
\nn \\
& + {\cal{O}}\left(\omega_{\rm BD}^{-2}, \delta e_{i-1}\right)\,,
\end{align}

Thus the burst algorithm in BD theory does properly map to the proposed ppE burst model. We find that
\begin{equation}
\alpha_{\rm ppE} = \beta_{\rm ppE} = 0\,,
\end{equation}
since there are no corrections to the orbital energy and angular momentum in BD theory. Furthermore, from the pericenter and eccentricity mapping, we see that
\begin{align}
\gamma_{\ppE} &= -\frac{5 \sqrt{2} \pi}{12} \eta \frac{{\cal{S}}_{\rm BD}^{2}}{\omega_{\rm BD}}\,,
\qquad
\bar{c}_{\ppE} = 3/2\,,
\\
\delta_{\ppE} &= \frac{6}{85} \frac{{\cal{S}}_{\rm BD}^{2}}{\omega_{\rm BD}} \,,
\qquad
\bar{d}_{\ppE} = -1\,,
\end{align}
which is very similar to the results found in EDGB. This is due to both theories having dipole radiation. 

\subsection{Projected Constraints}
We here present a back-of-the-envelope calculation aimed at providing an order-of-magnitude estimate of how well these eccentric burst algorithms could constrain modified gravity theories. We stress that these are not rigorous projected constraints in any way; true projections would require a full data analysis investigation using Bayesian tools, which we leave to future work. We consider the variations in the time and frequency mappings between boxes due to GR parameters and modified gravity coupling constants.

Let us begin by recalling that the mapping between time-frequency windows in GR depends only on the vector $\lambda_{\rm GR}^{a} = ({\cal{R}}, {\cal{M}}, e)$. Let us further assume that a sufficiently large number of bursts have been detected, such that in GR one could estimate these parameters from these burst observations. In a modified gravity theory, the mapping between time-frequency windows will not only depend on $\lambda_{\rm GR}^{a}$ but also on $\lambda_{\rm ppE}^{a} = (\alpha_{\rm ppE}, \beta_{\rm ppE}, \gamma_{\rm ppE}, \delta_{\rm ppE})$ as well as on $(\eta, {\cal{M}}, {\cal{R}})$ due to the $(M/r_p)$ dependence [see Eq.~\eqref{df-ppE}]. Therefore, a GW modification can be measured or constrained provided $\lambda_{\rm ppE}^{a}$ produced a modification to the time-frequency mapping that is \emph{at least} comparable to that introduced by $\lambda_{\rm GR}^{a}$, i.e.~
\begin{equation}
\label{variation}
\frac{\delta (\Delta t_{i,i-1}, \Delta f_{i,i-1})}{\delta \lambda_{\rm ppE}^{a}} \delta \lambda_{\rm ppE}^{a} > \left(\frac{\delta (\Delta t_{i,i-1}, \Delta f_{i,i-1})}{\delta \lambda_{\rm GR}^{a}}\right)_{\rm GR} \! \! \! \! \delta \lambda_{\rm GR}^{a}\,.
\end{equation}

Let us focus on the frequency mapping. One could also investigate GR deviation in the time mapping, which will depend on $\alpha_\ppE$ [Eq.~\eqref{eq:ppEexact1}], $\gamma_\ppE$ [Eq.~\eqref{rp-new}] and $\delta_\ppE$ [Eq.~\eqref{eq:deltas-new}]. Either way should lead to an order of magnitude estimate for the projected constraint that is comparable. The frequency mapping can be obtained by combining Eqs.~\eqref{f-new} and~\eqref{rp-new}:
\begin{align}
\Delta f_{i,i-1} &= \Delta f_{i,i-1}^{\rm GR} + \Delta f_{i,i-1}^{\rm ppE}\,,
\\
\Delta f_{i,i-1}^{\rm GR} &= \frac{23}{12} \frac{{\cal{M}}^{5/3}}{{\cal{R}}_{i-1}^{8/3}} \left(1 + \frac{121}{230} \delta e_{i-1}\right)\,,
\\
\label{df-ppE}
\Delta f_{i,i-1}^{\rm ppE} &= {\text{LO}}\left[\frac{23}{12} \beta_{\rm ppE} \eta^{-2/5\bar{b}} \frac{{\cal{M}}^{5/3+2/3\bar{b}_{\rm ppE}}}{{\cal{R}}_{i-1}^{8/3+2/3\bar{b}_{\rm ppE}}}
\right.\
\nonumber \\
& \left.
- \frac{3 \sqrt{2}}{4 \pi} \gamma_{\rm ppE} \eta^{-2/5\bar{c}_{\rm ppE}} \frac{{\cal{M}}^{2/3\bar{c}_{\rm ppE}}}{{\cal{R}}_{i-1}^{1+2/3\bar{c}_{\rm ppE}}}
\right.
\nonumber \\
& \left.
- \frac{85}{48} \delta_{\rm ppE} \eta^{-2/5\bar{d}_{\rm ppE}} \frac{{\cal{M}}^{5/3+2/3\bar{d_{\rm ppE}}}}{{\cal{R}}_{i-1}^{8/3+2/3\bar{d}}} \right]\,.
\end{align}
Using the values of the ppE parameters in BD and EDGB theory, we find,
\begin{align}
\Delta f_{i,i-1}^{\rm BD} &= \frac{1}{2} \frac{{\cal{S}}_{\rm BD}^{2}}{\omega_{\rm BD}} \frac{{\cal{M}} \eta^{2/5}}{{\cal{R}}^{2}}\,,
\\
\Delta f_{i,i-1}^{\rm EDGB} &= \frac{1}{4} {\cal{S}}_{\rm EDGB}^{2} \frac{{\cal{M}} \eta^{2/5}}{{\cal{R}}^{2}}
\end{align}
The form of these modifications is identical, since to leading PN order, both theories lead to a dipolar-type modification. 

With this, we can now evaluate projected constraints from Eq.~\eqref{variation}. This requires that we calculate the variation of $\Delta f_{i,i-1}$ with respect to the coupling constants of the theory. Recall that in BD the latter is $1/\omega_{\rm BD}$, while in EDGB it is just $\xi$. The evaluation of Eq.~\eqref{variation} also requires us to multiply this variation by $\delta \lambda^{a}_{\GR}$ and $\delta \lambda^{a}_{\ppE}$. The latter is simply $1/\omega_{\rm BD}$ in BD theory and $\xi$ in EDGB theory, because their values in GR are zero. 

We will consider the constraints that can be placed using the NSBH system considered in Appendix~\ref{validity}, as it will have the largest dipole contribution of the three systems studied there. This means that $m_{\rm NS} = 1.4 M_\odot$ and $m_{\rm BH} = 10 M_\odot$, which then automatically means that $\eta \approx 0.108$ and ${\cal{M}} = 1.23 M_\odot$. In BD, we choose $s_{1} = s_{\rm BH} = 0.5$ and $s_{2} = s_{\rm NS} = 0.2$; the sensitivity of neutron stars varies depending on their mass and equation of state, but the value $0.2$ is representative~\cite{Will:1989sk}. In EDGB, we set $\zeta_{2} = \zeta_{\rm NS} = 0$ since the corrections we have derived only apply to black holes. We will also need to choose values for the orbital eccentricity and the radius of curvature, which recall was defined as ${\cal{R}} = r_{p}^{3/2}/M^{1/2}$. We will here choose the initial value of $e$ and $r_p$ from Table I, to calculate $\delta e$ and ${\cal{R}}$.

In an optimal scenario, $\delta \lambda^{a}_{\GR}$ scales roughly as $1/\rho_{\rm GR}$, where $\rho_{\GR}$ is the signal-to-noise ratio in GR, defined as 
\begin{equation}
\label{snr-gr}
\rho_{\GR}^{2} = 4 \int_{0}^{\infty} \frac{|\tilde{h}_{\GR}(f)|^{2}}{S_{n}(f)} df\,,
\end{equation}
where $\tilde{h}_{\GR}(f)$ is the Fourier transform of the GW signal (i.e.~the frequency-domain waveform) and $S_{n}(f)$ is the spectral noise density of the detector considered. For our purposes, we will consider the Advanced LIGO detector, for which $S_{n}(f)$ is given e.g.~in Eq.~(2.1) of~\cite{mishra}. To obtain the $\tilde{h}_{\GR}(f)$, we use the method described in Appendix~\ref{validity} to obtain the evolution of the orbit under radiation reaction. We then compute the time-domain waveform using Eq.~(11) of~\cite{1977ApJ...216..610T} and use FFTW to obtain $\tilde{h}(f)$. For the NSBH system considered, we set the distance to the source to be 600 Mpc, for which we obtain $\rho_{\GR} = 13.2$, after approximately $3,650$ bursts. 

With this at hand, the right-hand side of Eq.~\eqref{variation} can be easily computed using the data from Table~\ref{ic}. In EDGB, the best constraint is  then
\begin{align}
\xi^{1/4} &\lesssim 3 \times 10^{3} \; {\rm m} \; \left(\frac{m_{\rm BH}}{10 M_{\odot}}\right) \left(\frac{{\cal{M}}}{1.23 M_{\odot}}\right)^{1/6} \left(\frac{\eta}{0.108}\right)^{-1/10} 
\nonumber \\
& \times \left(\frac{{\cal{R}}}{391 M}\right)^{-5/12} \left(\frac{\rho_{\rm GR}}{13.2}\right)^{-1/4} \,,
\end{align}
while in BD theory it is 
\begin{align}
\omega_{\rm BD} &\gtrsim 10^{3} \left(\frac{{\cal{S}}_{\rm BD}}{0.3}\right)^{2} \left(\frac{{\cal{M}}}{1.23 M_{\odot}}\right)^{-1/6} \left(\frac{\eta}{0.108}\right)^{1/10} 
\nonumber \\
& \times \left(\frac{{\cal{R}}}{391 M}\right)^{5/12} \left(\frac{\rho_{\rm GR}}{13.2}\right)^{1/4} \,.
\end{align}
In both cases, the best constraint comes from using $\delta \lambda^{a}_{\GR} = {\cal{R}}$. We should note that $\rho_{\rm GR}$ is not independent of the other parameters, namely $({\cal{R}}, {\cal{M}}, \eta)$. Thus, if one changes the SNR significantly from the above scaling, the other parameters must also be updated according to Eq.~\eqref{snr-gr}. These projected constraints are comparable to current constraints in the EDGB case, but not better than current constraints in the BD case. However, one should keep in mind that these projections are not rigorous but rather upper limits, since they do not account for parameter covariances.

\section{Discussion}
\label{conclusions}

We have proposed ppE deformations of the GW burst sequences describing highly eccentric compact object mergers, to allow for quantitative tests of dynamical, strong-field GR if detections of such events are made with an excess power stacking algorithm. The ppE model conveniently factors the mathematical description of the burst sequence into the GR prediction multiplied by polynomial functions of the inverse pericenter separation. These polynomials depend on eight parameters in total---four amplitude and four exponent coefficients---which should capture the leading order dynamics of a {\em class} of generic deviations from GR for non-spinning, highly eccentric mergers. We have explicitly shown this beginning from a low, PN order approximation to the GR equations, valid for large pericenter encounters, though posit that the same factorization will hold if the GR description is replaced by a more accurate model. Future work could investigate whether this is indeed the case, using higher order PN or EOB equations, or comparisons to numerical relativity results. It will also be necessary to include the effects of spin in the GR models. 

With the ppE model in hand, given detection of an event, or a population of events with the GR burst sequences, one can in principle quantitatively answer the following questions: How consistent are the signals with GR as the underlying theory governing the dynamics of the sources? If consistent with GR, what constraints can be placed on the coupling constants of specific alternative theories whose leading order deviations can be described by ppE parameters? If inconsistent with GR, what are the most likely ppE parameters, and what then does this tell us about the underlying theory? To see how effectively these questions could be answered in practice will require (for example) a Bayesian analysis as carried out in~\cite{cornishsampson,Sampson:2013lpa,Sampson:2013jpa} using ppE templates for quasi-circular inspiral. Issues that need to be addressed here include how well parameters (first GR, then ppE) can be extracted from a burst sequence with models of detector noise, and how accurately the base GR burst sequence needs to be modeled. It would also be interesting to see how much better, if at all, eccentric mergers are versus quasi-circular inspirals to test GR. 

Yet another interesting avenue for future research is to see whether a ppE-like approach could be adopted to extract finite size/matter effects from eccentric mergers where one or both members of the binary are neutron stars. In other words, the baseline ``GR'' component of the burst sequence would be the GR binary black hole model, while the ppE component would characterize tidal deformations, the effects of magnetic fields, the onset of disruption/merger, etc. One of the key pieces of physics governing neutron star structure, namely the equation of state at nuclear densities, is unknown, and properties such as the magnetic strength are expected to vary from one neutron star to the next, which in either case necessitates the introduction of additional parameters within the GR waveform model. If these effects can, at least to leading order, be captured by ppE deformations of vacuum mergers, they could automatically be searched for in a ppE analysis. Furthermore, this analysis would make transparent what the degeneracies between putative deviations from GR and finite body effects are.

\acknowledgments
This research was supported by NSF grants PHY-1065710 (FP), PHY-1305682 (FP), PHY-1114374 (NY), PHY-1250636 (NY), NASA grant NNX11AI49G, under sub-award 00001944 (FP and NY), and the Simons Foundation (FP). We would like to thank Katerina Chatziioannou, Neil Cornish, and Sean McWilliams for several discussions.

\appendix
\section{Validity of the Burst Algorithm}
\label{validity}

\begin{figure*}[thb]
\includegraphics[clip=true,height=8cm,width=8cm]{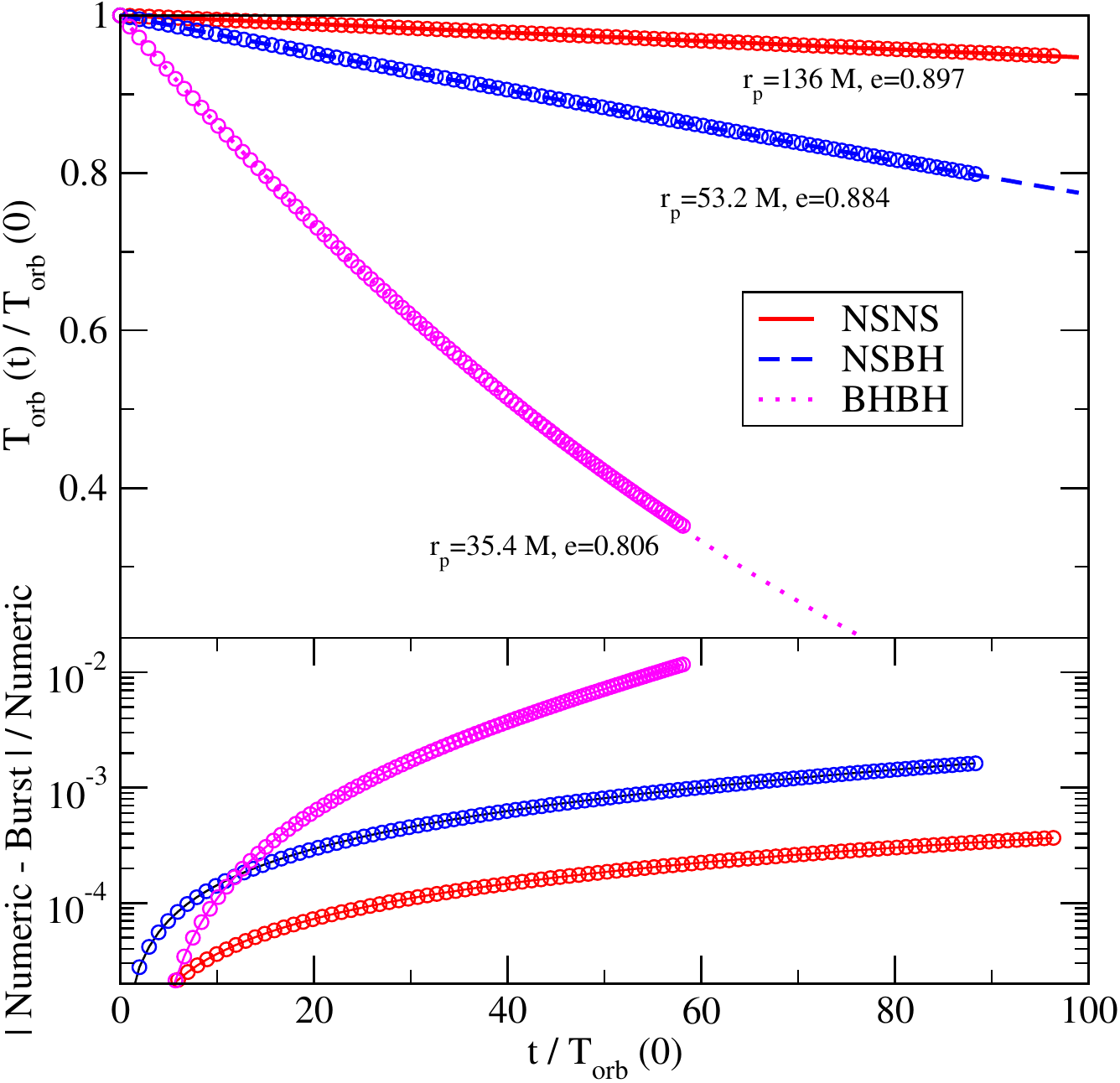}
\includegraphics[clip=true,height=8cm,width=8cm]{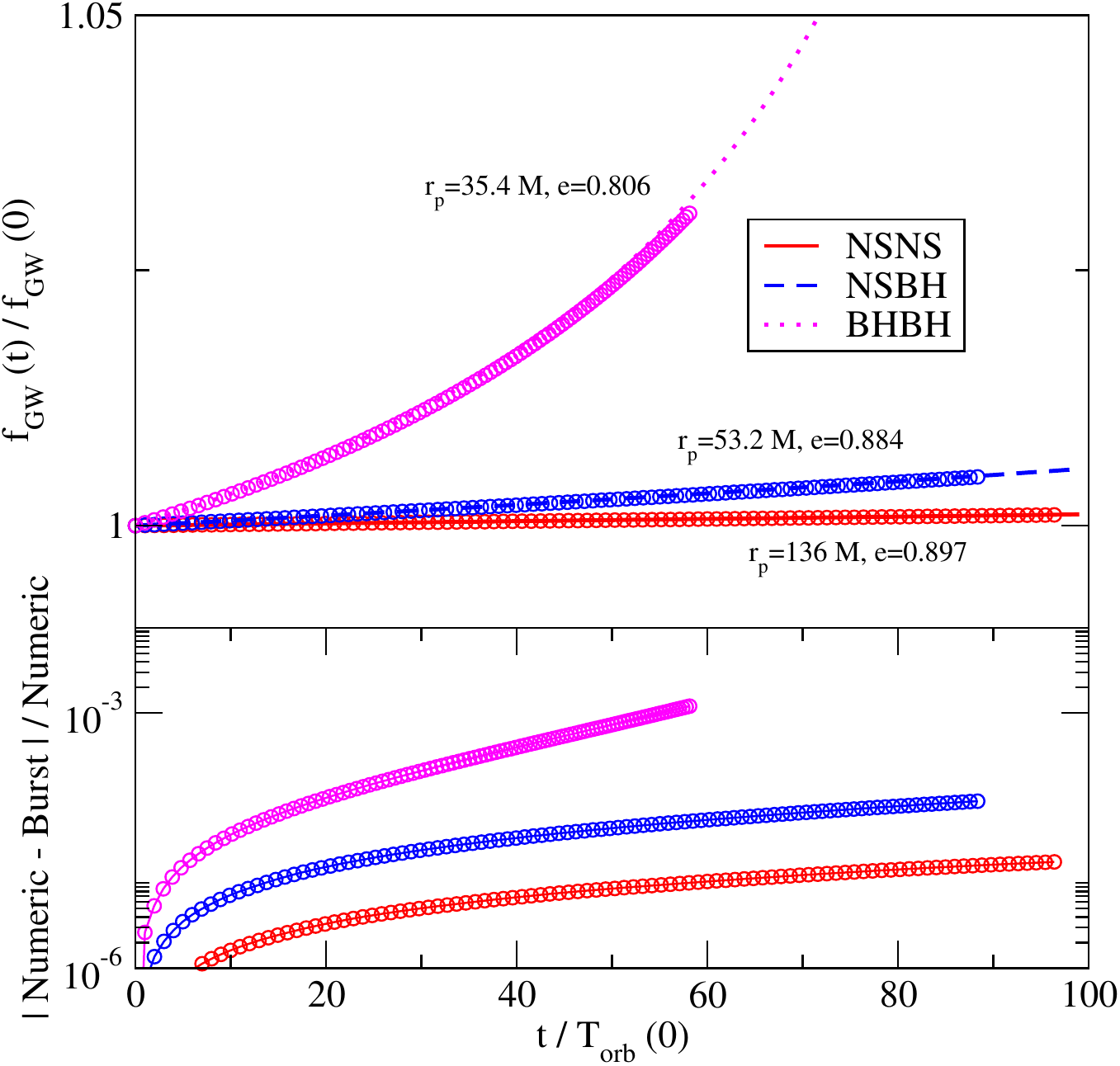}
\caption{\label{burst} Top panels: Orbital period and GW frequency as a function of time for the numerical solutions (lines) and the burst model approximation to it (circles). Three systems are shown: binary neutron star (red), mixed neutron star/black hole (blue), binary black hole (magenta). The values of $r_{p}$ and $e$ next to each line show their values after the last burst. Bottom panels: Relative fractional difference between the numerical solution and the burst model.}
\end{figure*}

The validity of the assumptions in our burst model simplification of the leading order PN dynamics---that all energy and angular momentum can be considered lost at pericenter passage, and between these episodes the orbits are Newtonian ellipses---can be gauged by comparing it to a numerical evolution of the full orbital dynamical equations with GW dissipation to leading PN order for an eccentric binary~\cite{Peters:1964zz,PetersMathews}. We consider three systems that could be detectable with current GW detectors: a NS binary, a mixed neutron star/black hole binary and a binary black hole, all at an initial separation corresponding to an initial GW frequency of $10$ Hz. The corresponding initial separations/eccentricities are such that these system would not have been formed via dynamical capture (see ~Eq.~\eqref{eq:rp-bound}), so could be examples of binaries formed via a 3 body interaction. For simply investigating the validity of the burst algorithm, the particular formation mechanism is not relevant; what matters is that $e_0$ is initially close to unity. The numerical evolution of the semi-major axis and eccentricity of the orbit are given in~\cite{Peters:1964zz}, which we solve with the initial data in Table~\ref{ic}. Figure~\ref{burst} compares the orbital period and burst frequency computed with the burst algorithm to the numerical solutions for the first one hundred bursts. Observe that even for the most compact system, the algorithm is an excellent approximation with relative fractional errors smaller than $1\%$.
{\renewcommand{\arraystretch}{1.2}
\begin{table}
\begin{centering}
\begin{tabular}{cccccc}
\hline
\hline
\noalign{\smallskip}
	System & $m_{1} [M_{\odot}]$ & $m_{2} [M_{\odot}]$ & $\tilde{r}_{p}(0)$ & e(0) & $\tilde{a}(0)$ \\ 
\hline
\noalign{\smallskip}
	NS-NS & 1.4 & 1.4 & 136  & 0.9 & 1360  \\ 
	NS-BH & 1.4 & 10 & 53.5  & 0.9 & 535  \\
	BH-BH & 10 & 10 & 36.8 & 0.9 & 368  \\ 
\noalign{\smallskip}	
\hline
\hline
\end{tabular}
\end{centering}
\caption{\label{ic} Initial dimensionless pericenter radius and semi-major axis for a set of compact binaries studied. Notice that the initial conditions here exceed the minimum radius allowed for systems created by dynamical capture, and thus, such binaries are assumed to form some other way, e.g.~via the Kozai-Lidov mechanism.}
\end{table}}

The burst algorithm becomes inaccurate when the eccentricity has decreased such that the $\delta e_{i-1} \ll 1$ expansions used in Eqs.~\eqref{rp2} and~\eqref{e2} become inaccurate. By comparing the enhancement factors $g_{1}(e)$ and $g_{2}(e)$ with their high eccentricity expansions, one finds that requiring $e > e_{\rm min} \approx 0.7$ is sufficient to guarantee the expansions remain reasonably accurate. When $M/r_{p}$ becomes large, the perturbative expansion of the change in the constants of the motion breaks down. One can estimate this error by investigating the 1PN modification to the energy and angular momentum flux in eccentric orbits~\cite{Arun:2009mc}.  For the systems we considered here, the 1PN terms are of order $0.1 \%$--$1\%$ relative to the leading-order Newtonian terms we include in our analysis. One could improve on our GR burst model by keeping higher-order terms in $\delta e$ and $M/r_{p}$, or by replacing our analytic model by a purely numerical one.  

\section{ppE Formalism of the Effective One-Body Problem in Newtonian Gravity}
\label{one-body}
We here derive corrections to various orbital quantities from two generic deformations of GR to leading, Newtonian order in a weak-field expansion. Let us consider polynomial ppE deformations to the potential energy of a binary system of the form
\begin{equation}
U = - \frac{\mu M}{r} \left[1 + \alpha' {\left(\frac{M}{r_{p}}\right)}^{\bar{a}'}\right] + \frac{1}{2} \mu r^2 {\dot{\phi}}^{2} \beta' {\left(\frac{M}{r_p}\right)}^{\bar{b}'}\,,
\end{equation}
where $r$ is the radial coordinate and $\phi$ is the azimuthal angle. From the Lagrangian $L = 1/2 \mu (\dot{r}^{2} + r^{2} \dot{\phi}^{2}) - U$, specialized to the equator, the two constants of motion (associated with stationarity and axisymmetry) are
\begin{align}
\label{com-E}
E &= \frac{1}{2} \mu {\dot{r}}^2 + \frac{{L}^2}{2 \mu r^2} \left[1 + 3 \beta' {\left(\frac{M}{r_p}\right)}^{\bar{b}'}\right] 
\nn \\
&- \frac{\mu M}{r} \left[1 + \alpha' {\left(\frac{M}{r_p}\right)}^{\bar{a}'}\right]\,,
\\
\label{com-L}
L_{z} &= \mu {r}^{2} \dot{\phi} \left[1 - \beta' {\left(\frac{M}{r_p}\right)}^{\bar{b}'}\right]\,,
\end{align}
which are simply the ppE-modified energy and (z-component of) angular momentum. 

Let us calculate the turning points of the orbit. Setting $\dot{r} = 0$ in Eq.~\eqref{com-E} leads to the turning point equation:
\begin{equation}
{r}^{2} + \frac{\mu M}{E} \left[1 + \alpha' {\left(\frac{M}{r_{p}}\right)}^{\bar{a}'}\right] r - \frac{{L}_{z}^{2}}{2 \mu E} \left[1 + 3 \beta' {\left(\frac{M}{r_p}\right)}^{\bar{b}'}\right] = 0\,,
\end{equation}
the roots of which are the turning points $r_{\pm}$. The semi-major axis and eccentricity of the orbit are then $a = ({r_{+} + r_{-}})/{2}$ and $e = ({r_{+} - r_{-}})/({r_{+} + r_{-}})$, and inserting $r_{\pm}$ in terms of $(E,L_{z})$ and inverting, we find
\begin{align}
\label{EppE}
E &= - \frac{\mu M}{2 a} \left[1 + \alpha' {\left(\frac{M}{r_{p}}\right)}^{\bar{a}'}\right]\,,
\\
\label{LppE}
L_{z} &=\mu \sqrt{M a (1 - e^2)} \left[1 + \frac{1}{2} \alpha' {\left(\frac{M}{r_p}\right)}^{\bar{a}'} - \frac{3}{2} \beta' {\left(\frac{M}{r_p}\right)}^{\bar{b}'}\right].
\end{align}

In order to relate this energy and angular momentum to Eqs.~\eqref{E-eq} and~\eqref{L-eq}, we make the mapping
\begin{align}
\label{alpha-map}
&\alpha' \rightarrow \alpha\,,
\qquad
\bar{a}' \rightarrow \bar{a}\,,
\qquad
-\frac{\mu M}{2 a} \rightarrow E_{\GR}\,,
\\
\label{beta-map}
&\beta \left(\frac{M}{r_p}\right)^{\bar{b}} = \frac{1}{2} \alpha' \left(\frac{M}{r_p}\right)^{\bar{a}'} - \frac{3}{2} \beta' \left(\frac{M}{r_p}\right)^{\bar{b}'}\,,
\nn \\
&\mu \sqrt{M a (1-e^{2})} \rightarrow L_{\GR}\,.
\end{align}
Of course, Eq.~\eqref{beta-map} makes sense only if $\bar{a}' = \bar{b}'$, as $\bar{b}$ should not depend on $M$ or $r_{p}$. However, even when $\bar{a}' \neq \bar{b}'$, this mapping still works to leading-order in a weak-field expansion, thus neglecting the term with higher powers of $(M/r_{p})$, ie.~the second term in Eq.~\eqref{beta-map} if $\bar{b}' > \bar{a}'$, or vice-versa. 

Let us now derive the ppE-correction to the orbital period. Due to the term proportional to $r^{2}$ in the potential, the orbits will not be perfect ellipses, but will instead precess. As a result, the orbital period, defined as the time between successive pericenter passes, can be obtained by integration of the orbital equation of motion. Combining Eqs.~\eqref{com-L} and \eqref{com-E}, we find
\begin{align}
\frac{d\phi}{du} = &\left\{\frac{2 \mu E}{L_{z}^{2}} \left[1 - 2 \alpha' \left(\frac{M}{r_p}\right)^{\bar{a}'}\right] 
\right.
\nonumber \\
& \left.
+ \frac{2 M \mu^{2}}{L_{z}^{2}} \left[1 + \alpha' \left(\frac{M}{r_p}\right)^{\bar{a}'}  - 2 \beta' \left(\frac{M}{r_p}\right)^{\bar{b}'}\right] u 
\right.
\nonumber \\
& \left.
- \left[1 + \beta' \left(\frac{M}{r_p}\right)^{\bar{b}'}\right] u^{2}\right\}^{-1/2}\,,
\end{align}
which can be integrated to give
\begin{equation}
\label{ppE-eom}
r(\phi) = \frac{a (1-e^{2})}{1 + e \;{\rm cos}\left\{\phi \left[1 + \frac{1}{2} \beta' \left(\frac{M}{r_p}\right)^{\bar{b}'}\right]\right\}}\,,
\end{equation}
where we have defined $u = 1/r$ and where we have used Eqs.~\eqref{EppE} and~\eqref{LppE} to map $(E, L)$ to $(a, e)$. We recognize this as the Newtonian equation for an elliptical orbit with a modification to $\phi$ which induces precession. Changing variables to
\begin{equation}
\psi = \phi \left[1 + \frac{1}{2} \beta' \left(\frac{M}{r_p}\right)^{\bar{b}'}\right]\,,
\end{equation}
pericenter passage occurs when $\psi = 2 n \pi$, where $n \in \mathbb{Z}$. The orbital period is then easily obtained as
\begin{align}
\label{Torb-ppE}
T_{\rm orb} &= \int_{0}^{2\pi} \frac{d\psi}{\dot{\psi}} = 2\pi \sqrt{\frac{a^{3}}{M}} \left[1 - \alpha \left(\frac{M}{r_p}\right)^{\bar{a}}\right]\,,
\end{align}
where we have mapped $(\alpha', \bar{a}')$ to $(\alpha, \bar{a})$ in the final expression.

Let us also calculate the ppE-corrections to the pericenter velocity. At any point in the orbit, the velocity is related to the energy by the trivial Newtonian relation $E = \frac{1}{2} \mu v^2 + U$. Using Eqs.~\eqref{EppE} and~\eqref{LppE} and solving for the velocity gives the ppE-modified \emph{Vis-Viva} equation:
\begin{align}
v &= \sqrt{\frac{2 M}{r} - \frac{M}{a}} \left[1 + \frac{1}{2} \alpha' {\left(\frac{M}{r_p}\right)}^{\bar{a}'} 
\right.
\nonumber \\
&\left.
- \frac{1}{2} \frac{a^2 (1 - e^2)}{r (2a - r)} \beta' {\left(\frac{M}{r_p}\right)}^{\bar{b}'}\right]
\end{align}
The term proportional to $\beta'$ has divergences when $r=0$ and $r=2a$. However, from Eq.~\eqref{ppE-eom}, we see that $r$ never takes these values for elliptical orbits. The velocity at pericenter is obtained by setting $r = r_p = a (1 - e)$:
\begin{equation}
\label{vppE}
v_p = \sqrt{\frac{M (1+e)}{r_p}} \left[1 + \frac{1}{3} \alpha \left(\frac{M}{r_p}\right)^{\bar{a}} + \frac{1}{3} \beta \left(\frac{M}{r_p}\right)^{\bar{b}}\right]\,,
\end{equation}
where we have used Eqs.~\eqref{alpha-map} and~\eqref{beta-map} to map to the ppE parameters. One can also obtain the same result by calculating $\dot{\phi}$ in terms of $a$ and $e$, and then evaluating the resulting expression at $r = r_{p}$.
\bibliography{master}
\end{document}